\documentclass[reprint,amsmath,amssymb,aps,prb,superscriptaddress]{revtex4-1}

\pdfoutput=1

\usepackage{hyperref,url}
\usepackage{amsmath}
\usepackage{amsfonts}
\usepackage[capitalise]{cleveref}
\usepackage{placeins}
\hypersetup{breaklinks,colorlinks=true,linkcolor=blue,citecolor=blue,urlcolor=blue,filecolor=blue}
\usepackage{graphicx}
\usepackage{dcolumn}
\usepackage{mhchem}

\newcommand\mat\mathbf

\newcommand{\COMMENTED}[1]{}

\newcommand{\insertnew}[1]{#1}

\begin{document}
\title{Constrained-Path Auxiliary-Field Quantum Monte Carlo for Coupled Electrons and Phonons%: Application to The One- and Two-Dimensional Holstein and Hubbard-Holstein Models
}
\author{Joonho Lee}
\email{jl5653@columbia.edu}
\affiliation{
Department of Chemistry, Columbia University, New York, New York 10027, USA
}
\author{Shiwei Zhang}
\email{szhang@flatironinstitute.org}
\affiliation{Center for Computational Quantum Physics, Flatiron Institute, New York, NY 10010, USA}
\affiliation{Department of Physics, College of William and Mary, Williamsburg, Virginia 23187, USA}
\author{David R. Reichman}
\email{drr2103@columbia.edu}
\affiliation{
Department of Chemistry, Columbia University, New York, New York 10027, USA
}

\begin{abstract}
We present an extension of  constrained-path auxiliary-field quantum Monte Carlo (CP-AFQMC) for the treatment of correlated electronic systems coupled to phonons. The algorithm follows the standard CP-AFQMC approach for description of the electronic degrees of freedom while phonons are described in first quantization and propagated via a diffusion Monte Carlo approach. Our method is tested on the one- and two-dimensional Holstein and Hubbard-Holstein models. With a simple semiclassical trial wavefunction, our approach is remarkably accurate for $\omega/(2\text{d}t\lambda) < 1$ for all parameters in the Holstein model considered in this study. In addition, we empirically show that the autocorrelation time scales as $1/\omega$ for $\omega/t \lesssim 1$, which is an improvement over the $1/\omega^2$ scaling of the conventional determinant quantum Monte Carlo algorithm. In the Hubbard-Holstein model, the accuracy of our algorithm is found to be consistent with that of standard CP-AFQMC for the Hubbard model when the Hubbard $U$ term dominates the physics of the model, and is nearly exact when the ground state is dominated by the electron-phonon coupling scale $\lambda$. The approach developed in this work should be valuable for understanding the complex physics arising from the interplay between electrons and phonons in both model lattice problems and {\it ab-initio} systems. 
\end{abstract}
\maketitle
%\onecolumngrid

\section{Introduction}\label{sec:intro}
%The coupled motion of electrons and phononsCP-AFQMC
%has been of great interest in 
%condensed-matter physics.
The coupling of electrons to nuclear lattice distortions is 
%often modeled by
%the Holstein model
%or the Su-Schrieffer-Heger (SSH) model.
responsible for myriad important physical phenomena in bulk materials.\cite{giustino2017electron} In particular, the thermodynamic and transport properties of solids are crucially influenced by electron-phonon (el-ph) interactions.
Perhaps the most spectacular consequence of el-ph interactions is the emergence of superconductivity as described by the Bardeen-Cooper-Schrieffer (BCS) theory.  Here, the el-ph interaction mediates an effective electron-electron (el-el) attraction which results in the Cooper pairing of electrons of opposite spin.\cite{mahan2013many} The BCS theory provides a quantitative framework for the description of conventional superconductivity such as that found at low temperatures in simple metals.

A simple microscopic picture is unfortunately not available for
unconventional superconductors such as the cuprates, whose critical temperature ($T_c$) can be above 90K at ambient pressure.\cite{cava1987bulk,wu1987superconductivity}
%The contemporary consensus is that
It is believed that
the el-ph interaction alone cannot give rise to these high $T_c$ values. %and may not even play any significant role
\cite{scalapino2012common}
However, experimental evidence exists which indicates that non-negligible el-ph interactions are present in these materials.\cite{Muller1990,Song1995,Alexandrov2000,Cuk2005,Khatami2008,Gunnarsson2008,VanHeumen2009,Gadermaier2010,He2018,Yang2019,Hu2019,Rosenstein2019,Grissonnanche2020,Rosenstein2020,Banerjee2020,Shneyder2020,Mishchenko2020,Sreedhar2020,Peng2020} 
It remains unclear what role el-ph interactions play in the cuprates and related materials, and if a potentially delicate interplay between el-el and el-ph interactions may influence their superconducting properties.

The canonical model Hamiltonian used to capture the physics of the cuprates
is the two-dimensional (2D) repulsive Hubbard model.\cite{hubbard1963electron}
The ground state of the hole-doped 2D Hubbard model has been 
thought %suspected 
to support d-wave superconductivity for many years.\cite{anderson1987resonating}
%Interestingly, a 
A recent joint numerical study
using two state-of-the-art approaches, density matrix renormalization group (DMRG)
and constrained path (CP) auxiliary-field quantum Monte Carlo (AFQMC),
%strongly suggests
indicates that
the ground state of the standard 2D repulsive Hubbard model with near-neighbor hopping
supports modulated phases (e.g. stripes) that are not superconducting over a range of repulsion strengths and doping levels\cite{qin2020absence} expected to describe the cuprates.  %If correct, this study implicates 
This suggests that features beyond those included in the simple Hubbard model, such as the effects of multiple bands, longer ranged Coulomb interactions, and/or the role of el-ph interactions, may be needed %as necessary 
to tip the balance of the ground state towards superconductivity for realistic values of doping levels and the magnitude of el-el repulsions.

Our work is motivated by precisely these considerations, namely the development of a scalable and accurate numerical approach that can treat el-ph effects on the same footing as el-el correlations.  This is a challenging task, as treating the complex electronic degrees of freedom in the pure 2D Hubbard model is already difficult, even with state-of-the-art numerical approaches.\cite{LeBlanc2015,zheng2017stripe,qin2020absence} The addition of el-ph effects, as contained in, e.g., the 2D Hubbard-Holstein model, thus requires non-trivial
extensions of these approaches in order to treat electrons and phonons on an equal footing.

Several methods have been formulated or extended to  coupled el-ph problems, including 
DMRG,\cite{jeckelmann1998density,jeckelmann1999metal,tezuka2005dmrg,tezuka2007phase,fehske2008metallicity,ejima2010dmrg}
variational exact diagonalization,\cite{bonvca1999holstein} variational Monte Carlo,\cite{ohgoe2014variational,ohgoe2017competition,karakuzu2017superconductivity}
 dynamical mean-field theory,\cite{Jeon2004,Paci2006,Werner2007,Murakami2013,Li2017} density matrix embedding theory, \cite{Sandhoefer2016,Reinhard2019}  \insertnew{and coupled-cluster theory.\cite{sibaev2020molecular,dresselhaus2020coupling,white2020coupled}}
There are difficulties facing each approach. For example, large el-ph couplings and/or small phonon frequencies are challenging to handle 
in most methods based on a second quantized representation of phonons, because of 
the necessity of truncating the phonon Hilbert space. When a  large number of phonons per site is required, the computational cost 
associated with treating them can grow prohibitively expensive. In addition to the demand of treating the phononic Hilbert space, there is of course the interacting many-electron problem. Clearly, the treatment of correlated el-ph coupled systems in two and higher dimensions over a wide range of the parameter space in an exact or near-exact manner is a forefront challenge.

The method that we propose here is an extension of the CP-AFQMC method developed and popularized by Zhang and co-workers.\cite{Zhang1995,Zhang1997}  For purely electronic problems, the CP-AFQMC %CP-AFQMC 
approach is similar to the determinant quantum Monte Carlo (DQMC) method\cite{Blankenbecler1981,Scalapino1981,Scalapino1981a,Johnston2013,Mendl2017,Karakuzu2018,Costa2020} in the sense that the two-body propagation is aided by the %discrete 
Hubbard-Stratonovich transformation %(or the Hirsch decomposition) 
\cite{Hirsch1983} and is formulated in the space of determinants. 
There, however, are several key differences. CP-AFQMC reformulates the imaginary-time propagation by working with open-ended random walks.  
An exact boundary condition is introduced in auxiliary-field space, which can be approximately imposed using a trial 
wave function, to avoid the notorious fermion sign problem. The open-ended random walk approach allows easy access to zero temperature results, 
and is often much less prone to ergodicity problems in the Monte Carlo sampling. 
%The AFQMC approaches introduces importance sampling which 
%The key differences are that CP-AFQMC introduces a constraint in the imaginary-time propagation by working with an open-ended random walk to eliminate the infamous fermion sign problem and works directly at zero temperature.
Moreover, CP-AFQMC can be naturally extended to {\it ab-initio} Hamiltonians while coping with the fermionic phase problem associated with these more complex models
using the phaseless approximation instead of the constrained path approximation. \cite{zhang2003quantum,al2006auxiliary}
%However, due to
Because of the constraint imposed on walker trajectories,
CP-AFQMC is no longer exact, unlike DQMC.  
Furthermore, due to the constraint, the ground-state energy computed via the usual mixed estimator %CP-AFQMC 
is not variational.\cite{carlson1999issues} On the other hand, CP-AFQMC can be used to access a wider range of interaction strengths and doping regimes in which DQMC cannot be used due to the inherent sign problem.  It should be noted that, in addition to its flexibility, CP-AFQMC %CP-AFQMC 
has been shown to yield excellent accuracy for strongly correlated electrons.\cite{zheng2017stripe,qin2020absence}

In this work we devise an extension of CP-AFQMC to treat both electrons and phonons on an equal footing, while retaining its %the benefits of CP-AFQMC 
benefits for electrons. Our framework is %most 
similar to the extension of Green's function Monte Carlo (GFMC) as formulated by McKenzie and others,\cite{mckenzie1996quantum} where the phonons are treated in a first quantized space.
We present the formulation of this new CP-AFQMC %CP-AFQMC 
approach, provide thorough benchmark results on the 1D and 2D Holstein and Hubbard-Holstein models for various phonon frequencies and el-ph couplings, and discuss the current scope and limitations of the proposed approach.  

The paper is organized as follows:  In \cref{sec:model} we outline the model we study and the important parameters that control its physics.  In \cref{sec:cpmc} we outline our algorithm.  \cref{sec:trial} is devoted to a discussion of trial wave functions. \cref{sec:csmp2} and \cref{sec:lfpt2} discuss distinct perturbative approaches to the problem outlined in \cref{sec:model}.  
\cref{sec:holstein} and \cref{sec:hh} present results for the Holstein and Hubbard-Holstein models, respectively.  \cref{sec:abinitio} discusses the extension of our approach to realistic {\em ab-initio} problems.  In \cref{sec:conclusions} we conclude. 
%the treatment of electrons and phonons on an equal footing 
%extending the scope of state-of-the-art approaches.
%In particular, the Holstein model
%has received attention
%because of its potential relevance
%in understanding the competition between
%superconductivity and charge-density-wave (CDW) in Se-based superconductors, 
%\ce{C60}, and perhaps even cuprates.

\section{Model}
\label{sec:model}
\subsection{The Hubbard-Holstein Hamiltonian}
Although the approach we outline is general, we focus on a paradigmatic model
of a correlated system coupled to phonons, namely the Hubbard-Holstein model.\cite{hubbard1963electron,Holstein1959}
The Hubbard-Holstein model is defined by the following Hamiltonian:
\begin{align}\label{eq:HH}
\hat{\mathcal{H}}
&=
\hat{\mathcal H}_\text{el}^{(1)} + \hat{\mathcal H}_\text{el}^\text{(2)} + \hat{\mathcal H}_\text{ph}
+ \hat{\mathcal H}_\text{el-ph},
\end{align}
where
\begin{align}
\hat{\mathcal H}_\text{el} & = 
\hat{\mathcal H}_\text{el}^\text{(1)} + \hat{\mathcal H}_\text{el}^\text{(2)},\\
\hat{\mathcal H}_\text{el}^\text{(1)} &= 
-t
\sum_{\sigma\in\{\uparrow,\downarrow\}}\sum_{\langle ij\rangle}
\hat{a}_{i_\sigma}^\dagger
\hat{a}_{j_\sigma},\\
\hat{\mathcal H}_\text{el}^\text{(2)} &= 
 U 
\sum_i
\hat{n}_{i_\uparrow}
\hat{n}_{i_\downarrow},\\
\hat{\mathcal H}_\text{ph}
&=
\omega \sum_i \hat{b}_i^\dagger \hat{b}_i,\\
\end{align}
and
\begin{align}
\hat{\mathcal H}_\text{el-ph}
&=
- g \sum_i \hat{n}_i (\hat{b}_i + \hat{b}_i^\dagger).
\end{align}
with
\begin{equation}
\hat{n}_i = \sum_{\sigma\in\{\uparrow,\downarrow\}}\hat{n}_{i_\sigma}
\end{equation}
%and $\alpha$ and $\beta$ denote the spin labels of electrons.
The nearest-neighbor electronic hopping is controlled by $t$ and the on-site repulsion is characterized by the parameter $U$. 
The phonons are treated as harmonic oscillators with a single frequency $\omega$. 
The electronic density is coupled to the phonon degrees of freedom characterized by a coupling constant $g$.

There are three relevant dimensionless parameters to define.
The first is the adiabaticity ratio in units of the hopping parameter
\begin{equation}
\alpha = \frac{\omega}t.
\end{equation}
The second is the effective on-site repulsion in units of the hopping parameter
\begin{equation}
\frac U t.
\end{equation}
Lastly, we define the dimensionless el-ph coupling $\lambda$,
\begin{equation}
\lambda = \frac {g^2} {2\text{d} t \omega},
\end{equation}
where d is the dimensionality of the system.
When $U$ is the dominant parameter, a spin density wave (SDW) phase similar to that found in the  Hubbard model is expected to arise.
When $\lambda$ dominates, a charge density wave (CDW) phase similar to that found in the Holstein model arises.
A metallic or superconducting phase can arise when the system transitions between these two phases.\cite{Costa2020}
%The existence of such an intermediate phase and the precise onset of the phase transitions depends on $\alpha$.
%\REMARKS{commented out sentence with dependence on $\alpha$. we're not sure that the phases exist right?}

\subsection{Phonons in First Quantization}
Since the Hamiltonian in \cref{eq:HH} does not commute with the phonon number operator $\hat{b}_{i}^\dagger\hat{b}_{i}$,
the number of phonons in the system is not conserved.
Therefore, one needs to work with an infinitely large phonon Hilbert space in order to study eigenstates of the Hubbard-Holstein model.
Methods working in a second quantized space such as DMRG\cite{jeckelmann1998density} generally require a specification of the maximum number of phonons {\it a priori} for the sake of computational tractability. Limiting the maximum number of phonons effectively truncates the infinite Hilbert space,
which may introduce significant errors, particularly when $\alpha$ is small and/or $\lambda$ is large.

For this reason we work within the framework of first quantization, namely with position and momentum operators on each site $i$,
\begin{align}
\hat{X}_i &= \sqrt{\frac{1}{2m\omega}} (\hat{b}_i^\dagger + \hat{b}_i),\\
\hat{P}_i &= i\sqrt{\frac{m\omega}{2}} (\hat{b}_i^\dagger - \hat{b}_i),
\end{align}
and thus re-express
\begin{align}
\hat{\mathcal H}_\text{ph} & = \sum_i (\frac{m\omega^2}{2} \hat{X}_i^2 + \frac{1}{2m} \hat{P}_i^2 - \frac\omega2),\\
\hat{\mathcal H}_\text{el-ph} & = -g\sqrt{2m\omega}\sum_i \hat{n}_i \hat{X}_i,    
\end{align}
in the Hubbard-Holstein Hamiltonian in \cref{eq:HH}.
\insertnew{Here, we introduced a fictitious mass $m$ and throughout this work we use $m = 1/\omega$.}
Working in a first quantized space allows one to work directly at the complete basis set limit for the phonons 
and avoids the issues posed by a truncated phonon Hilbert space.

\section{Constrained-Path Auxiliary-Field Quantum Monte Carlo}\label{sec:cpmc}

\COMMENTED{
Since electrons interact via the on-site repulsion $U$, the notorious fermionic sign problem arises in projector quantum Monte Carlo (QMC) approaches.
To remove the sign problem within the auxiliary-field QMC (AFQMC) approach, we resort to the constrained-path (CP) approximation\cite{Zhang1995,Zhang1997} proposed by Zhang, Carlson, and Gubernatis which introduces a systematic bias
depending on the quality of the trial wavefunction.  We refer to AFQMC with the CP approximation simply as CP-AFQMC.
}

AFQMC %CP-AFQMC 
for mixed fermions and bosons was first formulated and studied by Rubenstein, Zhang, and Reichman.\cite{rubenstein2012finite}
In their formulation, bosons are treated within a second-quantized framework. Therefore, their approach would naturally suffer from
the truncation of the infinite bosonic Hilbert space if applied to the Hubbard-Holstein model.
In this work, we will reformulate the procedure %CP-AFQMC
 to treat fermions in a second-quantized space and bosons in a first-quantized space. Such a formulation is closely related to that of Ref.~\citenum{mckenzie1996quantum}, however our work allows the control of the sign problem and introduces the full advantage of the CP-AFQMC approach in treating the 
 electronic degrees of freedom.
 % is the first to apply the CP approximation within this mixed-quantization framework.

In AFQMC, as in other projector QMC methods,\cite{blankenbecler1983projector} we obtain the ground state via
\begin{equation}
|\Psi_0\rangle \propto \lim_{\tau\rightarrow\infty} e^{-\tau \hat{\mathcal H}} | \Phi_0\rangle,
\label{eq:projection}
\end{equation}
where $|\Psi_0\rangle$ is the true ground state, $\tau$ denotes imaginary time, and $| \Phi_0\rangle$ is 
a trial wavefunction with non-zero overlap with the true ground state.
Since $\hat {\mathcal H}$ involves both fermions and bosons and so do the wavefunctions $|\Psi_0\rangle$ and $|\Phi_0\rangle$,
we represent these global vibronic  wavefunctions as a function of imaginary time $\tau$ in a mixed basis 
%by
%writing a global vibronic wavefunction (without importance sampling) at imaginary time $\tau$ as 
\begin{equation}
|\Psi(\tau)\rangle = \sum_{i} \omega_i |\psi_i(\tau),\mathbf X_i(\tau)\rangle,
\label{eq:wf-rep}
\end{equation}
%where each walker wavefunction 
where $|\psi_i\rangle$ is the electronic wavefunction and  $|\mathbf X_i\rangle$ is a set of coordinates that represents the phonon degrees of freedom. 
In our algorithm, these basis states each take a product form: 
%We also define
\begin{equation}
%|\psi_i(\tau)\rangle \otimes |\mathbf X_i(\tau)\rangle \equiv |\psi_i(\tau),\mathbf X_i(\tau)\rangle.
 |\psi_i(\tau),\mathbf X_i(\tau)\rangle \equiv |\psi_i(\tau)\rangle \otimes |\mathbf X_i(\tau)\rangle\,,
 \label{eq:prod-state-rep}
\end{equation}
%\REMARKS{I reordered above. OK?}
%We will assume that 
where $|\psi_i\rangle$ is a single Slater determinant.
We will show below that the projection process in Eq.~(\ref{eq:projection}) can be turned into a random walk in the space of product states 
of the form defined in Eq.~(\ref{eq:prod-state-rep}).
We note that it is also possible to work in momentum space ($|\mathbf P_i\rangle$),\cite{Hohenadler2004} however 
%in this study we choose to work in position space only.
it is more convenient to work in position space here since it makes the application of the e-ph coupling term straightforward.

%With this wavefunction form, the imaginary-time propagation is straightforward to perform.
We write the propagator for a finite timestep $\Delta \tau$ as
\begin{align}\nonumber
\exp(-\Delta \tau \hat{\mathcal H})
&=
e^{-\Delta \tau \hat{\mathcal H}_\text{el}^\text{(2)}}
e^{-\Delta \tau (\hat{\mathcal H}_\text{el}^{\text{(1)}}+\hat{\mathcal H}_\text{el-ph})}
e^{-\Delta \tau \hat{\mathcal H}_\text{ph}}\\
&+\mathcal O(\Delta \tau^2)
\end{align}
using the standard first-order Trotter approximation. 
By %the 
virtue of %the 
the Thouless theorem,\cite{thouless1960stability} $|\psi_i(\tau)\rangle$ remains a single Slater determinant after propagation via $\hat{\mathcal H}_\text{el}^\text{(1)}$ and $\hat{\mathcal H}_\text{el-ph}$ (note the latter is diagonal in $|\mathbf X_i\rangle$ space). 
$\hat{\mathcal H}_\text{el}^\text{(2)}$ is represented as a one-body operator coupled to Ising variables and therefore a single Slater determinant remains in the same manifold after propagation by $e^{-\Delta \tau \hat{\mathcal H}_\text{el}^\text{(2)}}$.
The phonon propagation generated by $\hat{\mathcal H}_\text{ph}$ follows a commonly used diffusion MC (DMC) algorithm.\cite{umrigar1993diffusion,hammond1994monte}

%We will elaborate on how the propagation within the Hubbard-Holstein Hamiltoninan is performed below. 

%In passing, we note that a higher-order Trotter approximation (i.e., the symmetric Trotter split) is possible,
%\begin{align}\nonumber
%\exp(-\Delta \tau \hat{\mathcal H})
%&=
%e^{-\frac{\Delta \tau}2 \hat{\mathcal H}_\text{ph}}
%e^{-\frac{\Delta \tau}2 (\hat{\mathcal H}_\text{el}^{\text{(1)}}+\hat{\mathcal H}_\text{el-ph})}\\\nonumber
%&\times e^{-\Delta \tau \hat{\mathcal H}_\text{el}^\text{(2)}}
%e^{-\frac{\Delta \tau}2 (\hat{\mathcal H}_\text{el}^{\text{(1)}}+\hat{\mathcal H}_\text{el-ph})}
%e^{-\frac{\Delta \tau}2 \hat{\mathcal H}_\text{ph}}\\
%& +\mathcal O(\Delta \tau^3).
%\end{align}
%While we use this symmetric for the numbers reported in this paper, we discuss implementation of the first-order propagator for simplicity.

%We will elaborate on how the propagation within the Hubbard-Holstein Hamiltoninan is performed below. 
Before elaborating on the propagation more concretely, let us introduce importance sampling, using a trial vibronic wavefunction $|\Psi_T\rangle$.
%We will make use of importance sampling via a trial vibronic wavefunction $|\Psi_T\rangle$, 
%and thus 
We re-write the global vibronic wavefunction in Eq.~(\ref{eq:wf-rep}) in the following form,  \cite{zhang2003quantum}
\begin{equation}
|\Psi(\tau)\rangle = \sum_{i} \omega_i 
\frac{| \psi_i(\tau), \mathbf X_i(\tau)\rangle}
{\langle \Psi_T | \psi_i(\tau), \mathbf X_i(\tau)\rangle}
\label{eq:walkers}
\end{equation}
to perform imaginary-time propagation, namely
\begin{equation}
|\Psi(\tau+\Delta\tau)\rangle
=
%e^{-\Delta \tau \hat{\mathcal H}_\text{el}}
%e^{-\Delta \tau \hat{\mathcal H}_\text{el-ph}}
%e^{-\Delta \tau \hat{\mathcal H}_\text{ph}}
e^{-\Delta \tau \hat{\mathcal H}_\text{el}^\text{(2)}}
e^{-\Delta \tau (\hat{\mathcal H}_\text{el}^{\text{(1)}}+\hat{\mathcal H}_\text{el-ph})}
e^{-\Delta \tau \hat{\mathcal H}_\text{ph}}
|\Psi(\tau)\rangle.
\end{equation}
%where
%$|\Psi_T\rangle$ and $|\Phi_T\rangle$ are electronic and vibrational trial wavefunctions, respectively.
%The phonon propagation is now then straightforward following the standard diffusion Monte Carlo (DMC) algorithm.
%Our goal is to sample
%the distribution of

In the propagation of the phonon degrees of freedom we sample from
the distribution 
\COMMENTED{
\begin{equation}
f (\mathbf X, \tau) 
= \langle \mathbf X | \Psi_T\rangle 
\langle \mathbf X | \Psi(\tau)\rangle 
= \Psi_T(\mathbf X) \Psi(\mathbf X, \tau).
%\Psi_T(\mathbf X) \Psi(\mathbf X).
\end{equation}
}
\begin{align}\nonumber
f [(\mathbf X(\tau+\Delta\tau)] 
&\equiv \frac
{\langle \Psi_T | \psi(\tau), \mathbf X(\tau+\Delta\tau) \rangle}
{\langle \Psi_T | \psi(\tau), \mathbf X(\tau) \rangle} \\
&
\times \langle X(\tau+\Delta\tau) | e^{-\Delta \tau \hat{\mathcal H}_\text{ph}} | X(\tau) \rangle\,,
\label{eq:ph-dmc}
\end{align}
where we have omitted the walker index $i$ in the subscript. 
%\REMARKS{Replaced $f$ formula above.  
%New formula OK?}
One can derive the following MC move for the updating of the variable $\mathbf X(\tau)\equiv \mathbf X$:
\begin{equation}
\mathbf X (\tau + \Delta \tau)
=
\mathbf X %(\tau)
+
\mathcal N (\mu = 0, \sigma = \sqrt{m\Delta\tau})
+
\frac{\Delta\tau}{m}
\frac{\nabla_{\mathbf X} \Psi_T(\mathbf X)}{\Psi_T(\mathbf X)}.
\end{equation}
Here $\mathcal N (\mu, \sigma)$ is a normally distributed random number with mean $\mu$ and variance $\sigma^2$,
and the last term is the so-called drift term. Updates for the walker weights are carried out as 
\begin{equation}
w (\tau + \Delta \tau)
=
w (\tau)
%\exp\left({-\frac{\Delta\tau}{2}(E_\text{loc}(\tau+\Delta\tau) + E_\text{loc}(\tau) - 2 E_\text{shift})}\right)
e^{{-\frac{\Delta\tau}{2}(E_\text{ph}(\tau+\Delta\tau) + E_\text{ph}(\tau) - 2 E_\text{shift})}},
\end{equation}
where
$E_\text{shift}$ is a constant shift that can be adjusted to control walker weight fluctuations, and we define
\begin{equation}
E_\text{ph} (\tau)
=
\frac{\langle \Psi_T | \hat{H}_\text{ph}|\mathbf X(\tau)\rangle}{\Psi_T(\mathbf X(\tau))}.
\end{equation}
This algorithm is the same as the standard diffusion Monte Carlo algorithm.\cite{moskowitz1982new,reynolds1982fixed,umrigar1993diffusion,hammond1994monte}

Propagation arising from $\hat{\mathcal H}_\text{el-ph}$ is straightforward to implement,
%The action of $\hat{\mathcal H}_\text{el-ph}$ on a walker wavefunction is trivial 
since
\begin{equation}
\sum_i \hat{n}_i \hat{X}_i  |\psi(\tau),\mathbf X(\tau)\rangle
= \sum_i \hat{n}_i X_i(\tau)  |\psi(\tau),\mathbf X(\tau)\rangle.
\end{equation}
$\sum_i \hat{n}_i X_i(\tau)$ is thus a diagonal matrix in the single-particle space with its $i$-th entry being $X_i(\tau)$.
%The application of the exponential of this diagonal matrix defines the action of $\hat{\mathcal H}_\text{el-ph}$ up to constant.
It is then straightforward to exponentiate this matrix and apply it along with $\hat{\mathcal H}_\text{el}^{(1)}$ to the Slater determinant.

%and compute the propagator in a matrix form.
Lastly, propagation generated by $\hat{\mathcal H}_\text{el}^{(2)}$ is the same as that for the standard %CP-AFQMC 
AFQMC algorithm for the Hubbard model.
%The one-body propagator due to $t$ is a simple matrix that can be applied to a walker determinant.
We employ the discrete Hirsch %-Fye 
spin decomposition for the two-body propagator \cite{Hirsch1983}:
%Namely, we represent the two-body propagator as a one-body operator coupled to an Ising variable
\begin{equation}
e^{-\Delta\tau U \hat{n}_{i_\uparrow}\hat{n}_{i_\downarrow}}
=
\frac12 e^{-\Delta\tau U (\hat{n}_{i_\uparrow}+\hat{n}_{i_\downarrow})/2}
\sum_{x_i=\pm1}e^{\gamma x_i (\hat{n}_{i_\uparrow} - \hat{n}_{i_\downarrow)}},
\label{eq:2eprop}
\end{equation}
where the constant $\gamma$ is determined by
\begin{equation}
\cosh(\gamma) = e^{-\Delta\tau U / 2}.
\end{equation}
For a given $\mathbf x$, the action of \cref{eq:2eprop} on a single Slater determinant keeps the Slater determinant in the single determinant manifold.
%The application of \cref{eq:2eprop} to walkers leads to the infamous fermionic sign problem.
%We remove the sign problem by employing the CP approximation.
In AFQMC %CP-AFQMC, 
we keep track of the overlap between the walker wavefunction and a chosen trial wavefunction.
More specifically, we measure the overlap ratio of the $i$-th walker,
%\REMARKS{
%I don't think this is correct: 
%\begin{equation}
%r_i = \frac
%{\langle \Psi_T | \psi_i(\tau+\Delta\tau), \mathbf X_i(\tau+\Delta\tau) \rangle}
%{\langle \Psi_T | \psi_i(\tau), \mathbf X_i(\tau) \rangle}.
%\label{eq:oratio}
%\end{equation}
%It should be:
\begin{equation}
r_i = \frac
{\langle \Psi_T | \psi_i(\tau+\Delta\tau), \mathbf X_i(\tau+\Delta\tau) \rangle}
{\langle \Psi_T | \psi_i(\tau), \mathbf X_i(\tau+\Delta\tau) \rangle}.
\label{eq:oratio}
\end{equation}
%(Even though the sampling of Ising fields is unchanged, the weight update is different.)
%Please confirm. And if the calculations actually used the first one, please check the results!
%}
If $r_i$ is negative, the constraint condition is invoked and %then 
we set the weight $w_i$ to zero, which then causes the walker to be  %subsequently the walker will be completely 
removed from the simulation. % via population control. 
Furthermore, we apply heat-bath sampling\cite{Zhang1997} using this ratio to importance sample the Ising variables.
This completes the description of our algorithm for the Hubbard-Holstein Hamiltonian.

%Lastly, t
The local energy evaluation at $\tau$ with the Hubbard-Holstein model is straightforward via the one-body walker Green's function
\begin{equation}
G_{i_\sigma j_\sigma}(\tau) \equiv 
\frac
{\langle \Psi_T | \hat{a}^\dagger_{i_\sigma} \hat{a}_{j_\sigma} |\psi(\tau),\mathbf X(\tau)\rangle}
{\langle \Psi_T | \psi(\tau),\mathbf X(\tau)\rangle},
\end{equation}
and the two-body walker Green's function,
\begin{equation}
\Gamma_{i_\uparrow i_\downarrow}
=
\frac
{\langle \Psi_T | \hat{a}^\dagger_{i_\uparrow} \hat{a}_{i_\uparrow} 
\hat{a}^\dagger_{i_\downarrow} \hat{a}_{i_\downarrow}
|\psi(\tau),\mathbf X(\tau)\rangle}
{\langle \Psi_T | \psi(\tau),\mathbf X(\tau)\rangle}.
\end{equation}
We will also need the mixed estimator for the phonon displacement,
\begin{equation}
\langle \hat{X}_i  \rangle (\tau)
\equiv
\frac
{\langle \Psi_T | \hat{X}_i |\psi(\tau),\mathbf X(\tau)\rangle}
{\langle \Psi_T | \psi(\tau),\mathbf X(\tau)\rangle}
=
X_i(\tau),
\end{equation}
and for the squared phonon momentum,
\begin{equation}
\langle \hat{P}_i ^2 \rangle (\tau)
%\equiv
%\frac
%{\langle \Psi_T | \hat{P}_i^2 |\psi(\tau),\mathbf X(\tau)\rangle}
%{\langle \Psi_T | \psi(\tau),\mathbf X(\tau)\rangle}
= -
\frac
{\langle \Psi_T | \nabla_{X_i}^2 |\psi(\tau),\mathbf X(\tau)\rangle}
{\langle \Psi_T | \psi(\tau),\mathbf X(\tau)\rangle}\,,
\label{eq:P2}
\end{equation}
where $\nabla_{X_i}^2$ can be applied to the left on $|\Psi_T\rangle$.
%We note that the evaluation of $\langle \hat{X}_i (\tau) \rangle$ is 
%trivial because
%we are already working with the eigenbasis of $\hat{X}_i$.
Using these mixed estimators, the local energy can be evaluated as
%{\color{red} Wick's theorem may not apply so the U term needs to be written in terms of two-body Green's function}
\begin{align}\nonumber
E_L
&=
-t
\sum_\sigma\sum_{\langle ij\rangle}
G_{i_\sigma j_\sigma}
+ U 
\sum_i
{\Gamma}_{i_\uparrow i_\downarrow}\\ \nonumber
%{G}_{i_\beta i_\beta}\\ \nonumber
&+
 \sum_i (\frac{m\omega^2}{2} X_i^2 + \frac{1}{2m} \langle \hat{P}_i^2\rangle - \frac\omega2)\\
& -g\sqrt{2m\omega}\sum_i (G_{i_\uparrow i_\uparrow} + G_{i_\downarrow i_\downarrow}) X_i.
\label{eq:eloc}
\end{align}

\section{Trial Wavefunctions}\label{sec:trial}

\COMMENTED{
In CP-AFQMC, the choice of a trial wavefunction is critical for achieving high accuracy.
In practice, 
it is also
important to make sure that
the form of trial wavefunction
does not pose any additional challenges in
efficiently performing CP-AFQMC.
both propagation with the CP approximation and local energy evaluation can be done efficiently without causing an exponentially costly step.
}
The choice of the trial wavefunction can affect the quality of the CP approximation in treating the electronic degree of freedom. 
It can also affect the computational efficiency in treating the electronic and especially the phononic degrees of freedom; in particular, a poor choice of the importance function can magnify or even introduce 
additional ergodicity issues, especially in an el-ph system when multiple phonon modes are pronounced. 
%In particular, i
It is highly advantageous if an accurate trial wavefunction allows  %the application of $e^{-\Delta\tau \hat{H}}$ to walkers, the evaluation of 
the overlap ratio in \cref{eq:oratio},
and %the evaluation of 
the local energy in \cref{eq:eloc}, to be efficiently evaluated. %performed.

\subsection{Semiclassical state}\label{ssec:semi}
The simplest variational trial wavefunction that we employ in this work takes
a simple product form between electronic and bosonic degrees of freedom,
\begin{equation}
|\Psi_T\rangle
=
|\psi_T\rangle \otimes |\phi_T\rangle,
\label{eq:cs}
\end{equation}
where $|\psi_T\rangle$ is a single determinant and
$|\phi_T\rangle$ is a coherent state (or a shifted harmonic oscillator state).
This wave function has been referred to as a ``semiclassical state'' in literature.\cite{kalosakas1998polaron,romero1998converging}
Due to its simple product form, there is no explicit entanglement between 
electrons and phonons.
The electronic trial wavefunction, $|\psi_T\rangle$, is parametrized by
orbital rotation %parameters 
$\pmb \theta$, 
\begin{equation}
|\psi_T(\pmb\theta)\rangle = e^{\hat{\kappa}}|\psi_0\rangle,
\label{eq:csel}
\end{equation}
where 
\begin{equation}
\hat{\kappa}
=
\sum_{ij}
(\theta_{ij} - \theta_{ji})
\hat{a}_i^\dagger
\hat{a}_j,
\label{eq:rot}
\end{equation}
and $|\psi_0\rangle$ is some initial determinant (normally obtained by diagonalizing the one-body electronic Hamiltonian).
%This simple single determinant trial wavefunction has been widely used in 
%the previous CP-AFQMC studies of the Hubbard model.\cite{Shi2013,LeBlanc2015,Qin2016,zheng2017stripe}
Single determinant trial wavefunctions have been widely used in 
previous AFQMC studies of the Hubbard model.\cite{Shi2013,LeBlanc2015,Qin2016,zheng2017stripe}

The phonon trial wavefunction, $|\phi_T\rangle$,
is parametrized by coherent state displacements $\pmb \beta$, 
\begin{equation}
\label{eq:csph}
|\phi_T(\pmb\beta)\rangle
=
e^{\sum_i
\beta_i (\hat{b}_i^\dagger - \hat{b}_i)}
|0\rangle
\equiv
\hat{\mathcal D}(\vec{\beta})
|0\rangle,
\end{equation}
where $\hat{\mathcal D}(\vec{\beta})$ is the displacement operator.
We optimize the energy of $|\Psi_T\rangle$ in \cref{eq:cs} variationally over
$\pmb \theta$ and $\pmb \beta$ and use this as the final trial wavefunction.
$|\phi_T(\pmb \beta)\rangle$ technically contains
infinitely many bosons, but it has 
a convenient property which allows for an efficient AFQMC algorithm %CP-AFQMC algorithm.
%Namely,
\begin{equation}
\hat{b}_i | \phi_T(\pmb \beta)\rangle
=
\beta_i | \phi_T(\pmb \beta)\rangle .
\label{eq:coherentidentity}
\end{equation}
Using this fact, one can show that
the projection of 
$\langle \phi_T(\pmb \beta)|$ on to $|\mathbf X \rangle$ is %simple, namely
\begin{equation}
\langle \phi_T  (\pmb \beta)| \mathbf X \rangle
=
\prod_i
\left(
\frac{m\omega}{\pi}
\right)^{\frac14}
%e^{-\frac{m\omega}2(X_i-\sqrt{\frac{2}{m\omega}}\beta_i)}.
e^{-\frac{m\omega}2(X_i-\sqrt{\frac{2}{m\omega}}\beta_i)^2}.
\label{eq:trialprojection}
\end{equation}
Similarly, the numerator of \cref{eq:P2} is straightforward to evaluate as well using
\begin{equation}
-\langle \phi_T  (\pmb \beta)| \nabla_{X_i}^2|\mathbf X \rangle =
-\nabla_{X_i}^2\langle \phi_T  (\pmb \beta)| \mathbf X \rangle.
\end{equation}
This semiclassical trial wavefunction therefore can be efficiently combined with the AFQMC algorithm.
%without any exponentially costly steps.

The variational energy of the semiclassical state can be 
obtained within the Born-Oppenheimer (BO) approximation.  After some algebra, it can be shown that the lowest energy of a semiclassical state can be obtained by minimizing 
\begin{equation}
\langle \hat{H}_\text{el} \rangle
-\frac{g^2}{\omega}
\langle
\sum_i \hat{n}_i^2
\rangle
\end{equation}
over the variational parameters in $|\psi_T\rangle$. For a fixed $\lambda$, variations in $\alpha$ do not change the energy of the semiclassical state.
%The energy dependence with respect to $\alpha$ with a fixed $\lambda$ is therefore solely due to the entanglement between 
%electronic and phonon degrees of freedom which is absent in this trial wavefunction.

%\REMARKS{consolidated discussions below - } 
The semiclassical state is exact in
{ (1) the limit $g\rightarrow \infty$, (2) the adiabatic limit $\omega \rightarrow 0$ (for a fixed $\lambda$) with $U \rightarrow 0$, and (3) the atomic limit $U \rightarrow \infty$.}
%We define the ``quantumness'' of phonons as\cite{proville2000small}
%\begin{equation}
%\zeta = \frac{\omega^2}{g^2} = \frac{\omega}{2dt\lambda}.
%\end{equation}
%Due to its classical nature, the wave function \cref{eq:cs}  %trial 
%becomes qualitatively inaccurate when the phonon frequency becomes large, inducing quantal nuclear effects.\cite{proville2000small} 
\insertnew{When $\lambda \rightarrow \infty$ (or $g \rightarrow \infty$ for a fixed $\omega$), the use of a single semiclassical state is not problematic even though the BO potential strongly bifurcates.  
This situation is physically similar to that of the atomic limit of the Hubbard model ($U\rightarrow \infty$)
where spin flips do not cause an energy penalty and a degeneracy occurs amongst all possible $2^N$ spin flips.
Similarly, in the Holstein model, charge swapping does not cause an energy penalty and the same macroscopic degeneracy occurs. 
%As a consequence, one can reliably perform CP-AFQMC with a single semiclassical state without suffering from an ergodicity problem. 
In other words, any one of the degenerate semiclassical states is equally well-suited as an approximate wave function.}

\insertnew{
Aside from these limits, the semiclassical state itself can be inaccurate, but we find that the subsequent AFQMC calculation with the semiclassical trial wave function
is often numerically exact.
The most difficult parameter regime for our AFQMC framework is when the Holstein coupling strength $g$ takes an intermediate value.
That is, $g$ is large enough that the el-ph correlation is strong but is small enough that the 
macroscopic degeneracy does not occur.
A straightforward way to probe these situations is to increase $\omega$ for a fixed $\lambda$ value.
In this case, $g$ can be much larger than $t$ but is always smaller than $\omega$ as long as $2\text{d}t\lambda < \omega$. %Therefore, the large $g$ limit can never be obtained this way due to the presence of even larger $\omega$.
In these situations, the semiclassical state can be a poor choice of a trial wave function in AFQMC, as we shall see.
This is 
because no %due to the lack of 
correlation between electronic and bosonic degrees of freedom 
is built into this trial wave function.}
%in the sampling procedure.}

\insertnew{
From a different point of view, the difficulty of semiclassical states was noted in the work of Proville and Aubry, who defined the ``quantumness'' of the phonons as \cite{proville2000small}
\begin{equation}
\zeta = \frac{\omega^2}{g^2} = \frac{\omega}{2\text{d}t\lambda} .
\end{equation}
As $\zeta$ increases, the semiclassical state qualitatively fails.\cite{proville2000small}
This is consistent with the picture described above in that
for a given $\lambda$, both $g$ and $\zeta$ increase as $\omega$ increases.
% We attribute these difficulties associated with the semiclassical states to the increase in the correlation between electrons and phonons instead of the quantal effect in phonons alone.
 We attribute the difficulties associated with semiclassical states to the increase in correlation between electrons and phonons instead of the quantal effect associated with the phonons alone.
}

\COMMENTED{
Lastly, we comment on the scope of this trial wavefunction. 
Our semiclassical state is exact for:
{ (1) the limit $g\rightarrow \infty$, (2) the adiabatic limit $\omega \rightarrow 0$ (for a fixed $\lambda$) with $U \rightarrow 0$, and (3) the atomic limit $U \rightarrow \infty$.}
Due to its classical nature, this wavefunction %trial 
becomes qualitatively inaccurate when the phonon frequency becomes large, inducing quantal nuclear effects. 
However, if $g\gg \omega$, the phononic behavior has classical features that render this simple %trial 
wavefunction qualitatively accurate.
We define the ``quantumness'' of phonons as
\begin{equation}
\zeta = \frac{\omega^2}{g^2} = \frac{\omega}{2\text{d}t\lambda}.
\end{equation}
%When combined with CP-AFQMC, the scope of this trial wavefunction becomes much broader but is not strictly quantitatively accurate in certain parameter regimes ($\zeta \gg1$), as we shall see.
This wave function can cause difficulties in AFQMC in regimes where it is very inaccurate  ($\zeta \gg1$), as we further discuss below.
%
%Another way to look at the scope of this trial wavefunction is to look at the transformed Hamiltonian via a unitary transformation defined by the displacement operators in \cref{eq:csph}:
%\begin{equation}
%\hat{\bar{H}} = \hat{H}_\text{el}
%\end{equation}
}

\subsection{Multivibronic state}
A linear combination of multiple semiclassical states can be used to
correlate electrons and phonons
\begin{equation}
|\Psi_\text{MS}\rangle
=
\sum_i
c_i
|\psi_T(\pmb\theta_i)\rangle
\otimes
|\phi_T(\pmb\beta_i)\rangle,
\end{equation}
where one may determine $\mathbf c$, $\pmb \theta$, and $\pmb \beta$ variationally.
We refer to this wavefunction as a multivibronic (MV) wavefunction.
Similar to the multi-Slater determinant trial states employed to study purely electronic systems.\cite{Shi2013} 
The MV wavefunctions of this form % are ultimately limited by an exponential wall since one 
would need 
exponentially many states for large systems. 
Nevertheless, due to their simplicity, multivibronic states can be valuable for the study of small systems.

%\onecolumngrid
\begin{figure*}
\includegraphics[scale=0.8]{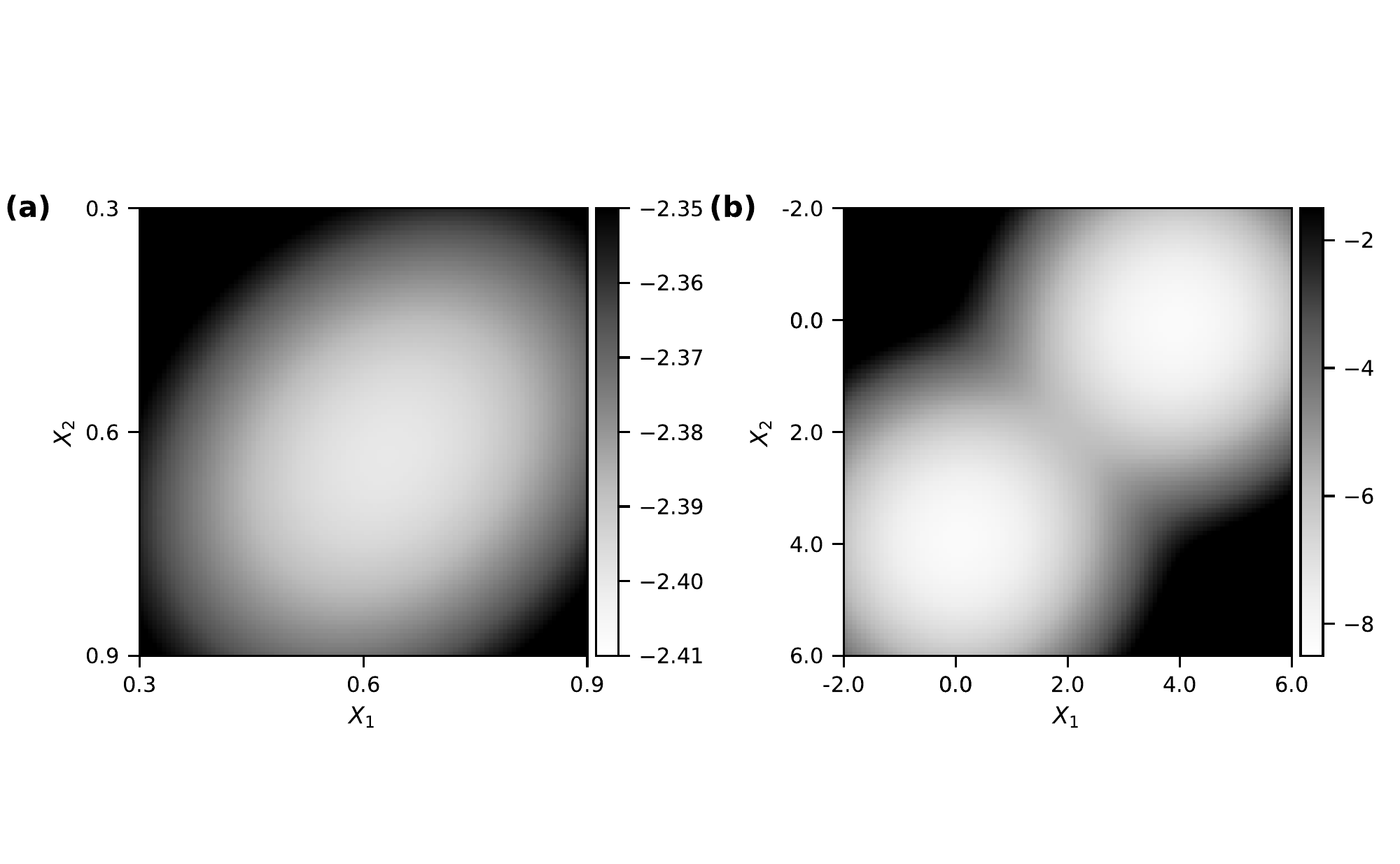}
\caption{\label{fig:hbo}
Born-Oppenheimer potential energy surfaces in units of $t$ for the 2-electron and 2-site Holstein model: (a) $\omega=t$, $\lambda = 0.1t$, and $g=0.447t$ and (b) $\omega=t$, $\lambda = t$, and $g=1.414t$.
The minimum in (a) is $E=-2.40 t$ at ($X_1$=0.63, $X_2$=0.63)
while
the two minima in (b) are $E=-8.25 t$ at ($X_1$=3.90, $X_2$=0.10) and ($X_1$=0.10, $X_2$=3.90), respectively.
}
\end{figure*}
%\twocolumngrid

A particular flavor of MV wavefunction that we focus on in this work is closely tied to the underlying order encoded in the semiclassical states themselves.
Let us consider the two-electron, two-site Holstein model.
It is well known that for the Holstein model
at a large coupling $\lambda$, the BO surface bifurcates and
develops into a double well potential.\cite{Holstein1959,Holstein1959a}
For the two-site problem, the BO potential energy surfaces (PESs) are
characterized by
\begin{align}\nonumber
\hat{H}_\text{BO}(X_1, X_2) &= \hat{H}_\text{el} -g\sqrt{2m\omega}(\hat{n}_1X_1 + \hat{n}_2X_2)\\
&+ \frac{m\omega^2}{2} ( X_1^2 + X_2^2),
\label{eq:hbo}
\end{align}
where $X_1$ and $X_2$ are constant scalars denoting the coordinates of the classical phonons. 
We can find the ground-state electronic wavefunction of the Hamiltonian in \cref{eq:hbo} by exactly diagonalizing it and forming a potential energy surface for each combination of ($X_1$,$X_2$).
In \cref{fig:hbo}, a representative example of the BO PESs is given. \cref{fig:hbo}(a) illustrates an example of the weak coupling case,
%. $\lambda$ is so small that 
where the coherent states have the same centers for all sites and no charge modulation occurs. In such cases, the minimum BO state (i.e. the semiclassical state) is an excellent variational %trial 
wavefunction. % for this problem.
In \cref{fig:hbo}(b) there are two distinct minima with equal BO energies. 
Here, a wave function of a single semiclassical state with a Gaussian function centered at one of the two BO minima in position space
would not provide a good description of the system. When used as an importance function, it can introduce or exacerbate ergodicity problems in the 
Monte Carlo sampling and induce a large or even infinite variance in the energy fluctuations.
%\REMARKS{general comment - I think there are 2 separate issues: one is how good/bad a variational wf is as a description of the system (e.g.
%giving only one well when the actual system has two); the other is whether the QMC sampling itself might have difficulty when the landscape 
%has double-well-like features. The two can become entangled when we use (1) as a trial wf for (2), but conceptually it's important to separate them
%wherever possible.
%This applies to the suggestions in Sec C below as well}

%\COMMENTED{
%In the intermediate coupling regime where $\lambda \sim t$, proper sampling of the 
%the CP-AFQMC walkers requires the exploration of
%both minima. 
%However, if a walker starts in one minimum,
%importance sampling may not enable the sampling of the other minimum, creating an ergodicity problem.
%%This problem is most severe when considering minima separated by a large distance.
%The phonon wavefunction in the semiclassical state is a gaussian function centered at one of the two BO minima in position space. 
%Therefore, the walker overlap with the trial wavefunction will decay exponentially as the walker moves away from its center.
%If two minima are well separated, the walker overlap will vanish before the walker finds another minimum.
%}

%\subsubsection{Thouless path trial wavefunction}
%To cope with this ergodicity problem, 
We propose the following %trial 
improved variational wavefunction in this situation for the Holstein model on a bipartite lattice.
At half-filling there are two {\it exactly} degenerate semiclassical states.
In particular, one state is characterized by
\begin{equation}
\beta_i^{(1)} = 
\begin{cases}
\beta_e , & \text{if}\ i \ \text{on A sublattice}\\ %\text{if}\ i \ \text{mod}\ 2 = 0\\
\beta_o , &  \text{if}\ i \ \text{on B sublattice}, % \text{if}\ i \ \text{mod}\ 2 = 1,
\end{cases}
\end{equation}
%\REMARKS{OK? (even and odd sites do not generalize to 2D as cleanly - depends on how one label the sites)}
where $i$ is a site index.
%It is trivial to find another degenerate solution that is characterized by
The other degenerate solution is given by switching the A and B sublattices.
\COMMENTED{
\begin{equation}
\beta_i^{(2)} = 
\begin{cases}
\beta_o , & \text{if}\ i \ \text{mod}\ 2 = 0\\
\beta_e , & \text{if}\ i \ \text{mod}\ 2 = 1.
\end{cases}
\end{equation}
}
One can smoothly interpolate between the two states by defining a convex combination
\begin{equation}
\vec{\beta}(\alpha) = \alpha \vec{\beta}^{(1)} + (1-\alpha) \vec{\beta}^{(2)}
\label{eq:interp}
\end{equation}
for $\alpha \in [0,1]$.
For each $\vec{\beta}(\alpha)$, we find a single determinant that minimizes the 
energy of a single semiclassical state.
One can take a linear combination of all of these semiclassical states along the line that interpolates two solutions
to form a MV %trial 
wavefunction.
We refer this to as the Thouless path (TP) %trial 
wavefunction, $|\Psi_\text{TP}\rangle$,
\begin{equation}
|\Psi_\text{TP}\rangle = 
\sum_\alpha
c_\alpha |\Psi_T(\beta(\alpha))\rangle,
\label{eq:tp}
\end{equation}
where $c_\alpha$ is determined by variationally minimizing the energy.
The construction of the TP %trial 
wavefunction can be generalized to arbitrary filling fractions and number of sites
because different filling fractions
simply give rise to ordered states
with different wavelengths. 
%Most importantly, the construction of the TP trial wavefunction and its use in CP-AFQMC does not scale exponentially in system size.
The cost for its construction is negligible compared to the optimization of a semiclassical state.
Its use in AFQMC as a trial wave function simply
introduces a prefactor depending on the number of states included in \cref{eq:tp}.
We will refer to a TP wave function with $n$ interpolation values of $\alpha$ as TP($n$).  
\insertnew{While TP wave functions provide a simple and accurate importance %sampling 
function for bifurcated potential energy surfaces,
they also become inaccurate when the correlation between electrons and phonons becomes strong. }
%Therefore, there is a need for a trial wavefunction
%which encodes the el-ph correlation more directly (i.e., beyond a linear combination of semiclassical states) such that the subsequent AFQMC importance sampling is accurate and efficient.}
%In the following, we will refer to CP-AFQMC calculations with a TP trial wavefunction as CP-AFQMC($n$) with $n$ noting the total number of semiclassical states.

\COMMENTED{
When $\lambda \rightarrow \infty$, the use of a single semiclassical state is no longer problematic even when the BO potential strongly bifurcates.  
%In particular, t
This situation is physically similar to that of the atomic limit of the Hubbard model ($U\rightarrow \infty$)
where spin flips do not cause an energy penalty and a degeneracy occurs amonst all possible $2^N$ spin flips.
Similarly, in the Holstein model, charge swapping does not cause an energy penalty and the same macroscopic degeneracy occurs. 
%As a consequence, one can reliably perform CP-AFQMC with a single semiclassical state without suffering from an ergodicity problem. 
In other words, any one of the degenerate semiclassical states is equally well-suited as an approximate wave function.
%as a trial wavefunction for CP-AFQMC.
}
\subsection{Variational Lang-Firsov trial wavefunctions}\label{ssec:vlf}

%Besides the MV state we have discussed, a simple way
A simple, widely used way
to incorporate
correlation effects between electrons and phonons is
to use the polaron transformation or the Lang-Firsov (LF) transformation,\cite{lang1963kinetic}
\begin{equation}
|\Psi_\text{LF}\rangle
=
\hat{U}_\text{LF}
(\pmb \xi)
|\psi_T(\pmb\theta)\rangle
\otimes
|\phi_T(\pmb\beta)\rangle,
\label{eq:lfstate}
\end{equation}
where
\begin{equation}
\hat{U}_\text{LF}(\pmb \xi)
=
e^{\frac{1}{\sqrt{2}}\sum_i \xi_i\hat{n}_i(\hat{b}^\dagger_i - \hat{b}_i)} ,
\label{eq:lfU}
\end{equation}
and the set $\pmb {\xi}$ are referred to as the LF %Lang-Firsov 
amplitudes which are variational parameters along with $\pmb \theta$ and $\pmb \beta$.
\COMMENTED{ % moved to further below:
This unitary transformation can be thought of as a simple Jastrow factor that
encodes correlation between the electronic density and the phonon momentum on a site.
%\insertnew{However, 
%we note that unlike Jastrow factors (which are not unitary)
However, $\hat{U}_\text{LF}(\pmb \xi)$ is unitary and we thus expect this transformation to behave differently from %compared to the use of 
Jastrow factors in el-ph problems.\cite{ohgoe2014variational,ohgoe2017competition,karakuzu2017superconductivity} 
It is also different from the coupled-cluster operators considered in recent studies of el-ph problems as well.\cite{sibaev2020molecular,dresselhaus2020coupling,white2020coupled}
}From the wavefunction viewpoint, \cref{eq:lfstate} provides
a way to explicitly build a wavefunction with non-perturbative el-ph correlation on top of semiclassical states via a unitary transformation.
%Lastly, t
Typical LF implementations involve the phonon vacuum state as opposed to the coherent state in \cref{eq:lfstate}. We find %found 
that having the coherent state provides additional variational flexibility and thereby yields lower energies compared to those that use the vacuum state. Since it does not complicate the underlying optimization problem, we use the coherent state as written in \cref{eq:lfstate}.

%The most naive implementation of the LF form as a trial wave function in CP-AFQMC does not scale well.
%We will leave the exploration of those for a future study, but will 
%briefly review second-order perturbation theory based on the LF state below. 

\insertnew{While the details of the LF %Lang-Firsov 
transformation and its % the 
variational optimization have been well-documented,\cite{lang1963kinetic,silbey1984variational,harris1985variational,pouthier2013reduced} we briefly summarize them to provide a self-contained description.
Our goal is to simultaneously determine $\pmb \xi$, $\pmb \theta$, and $\pmb \beta$ variationally.
To carry this out, we find that it is simpler to work with the unitary-transformed Hamiltonian, $\hat{H}^\text{LF}$, based on $\hat{U}_\text{LF}$,
and optimize the variational energy of $\hat{H}^\text{LF}$ evaluated with the semiclassical wavefunction.}
We start from %the fact that 
\begin{equation}
\hat{U}_\text{LF}(\pmb \xi)^\dagger
a_{i_\sigma}^\dagger
a_{j_\sigma}
\hat{U}_\text{LF}(\pmb \xi)
=
a_{i_\sigma}^\dagger
a_{j_\sigma}
e^{(-\xi_i(b^\dagger_i-b_i)+ \xi_j(b^\dagger_j-b_j))}\,
\end{equation}
and
\begin{align}
%\hat{U}_\text{LF}(\pmb \xi)^\dagger a_{i_\sigma}^\dagger a_{j_\sigma}
%\hat{U}_\text{LF}(\pmb \xi)
%&=
%a_{i_\sigma}^\dagger
%a_{j_\sigma}
%e^{(-\xi_i(b^\dagger_i-b_i)+ \xi_j(b^\dagger_j-b_j))},\\
\hat{U}_\text{LF}(\pmb{\xi})^\dagger b_i^\dagger \hat{U}(\pmb \xi) &= b_i^\dagger - \xi_i\hat{n}_i,\\
\hat{U}_\text{LF}(\pmb{\xi})^\dagger b_i \hat{U}(\pmb \xi) &= b_i - \xi_i\hat{n}_i.
\end{align}
The LF transformed Hamiltonian reads
\begin{align}\nonumber
\hat{{\mathcal H}}_\text{el}^\text{LF}
&=
-t \sum_\sigma \sum_{\langle ij \rangle} 
a_{i_\sigma}^\dagger
a_{j_\sigma}
e^{(-\frac{\xi_i}{\sqrt{2}}(b^\dagger_i-b_i) +\frac{\xi_j}{{\sqrt{2}}}(b^\dagger_j-b_j))}\\ \label{eq:lfel}
&+
U\sum_i
\hat{n}_{i_\uparrow}
\hat{n}_{i_\downarrow},\\
\hat{{\mathcal H}}_\text{ph}^\text{LF}
&=
\omega
\sum_i(\hat{b}^\dagger_i + \frac{\xi_i}{\sqrt{2}}\hat{n}_i)(\hat{b}_i + \frac{\xi_i}{\sqrt{2}}\hat{n}_i), \\ 
\end{align}
and
\begin{align}
\hat{{\mathcal H}}_\text{el-ph}^\text{LF}
&=
-g
\sum_i\hat{n}_i(\hat{b}_i + \hat{b}_i ^\dagger + \sqrt{2}\xi_i\hat{n}_i ).
\end{align}
\insertnew{All of the energy terms are straightforward to evaluate with semiclassical trial wavefunctions. The electronic kinetic energy
is more complex than its bare Hamiltonian counterpart due to the presence of exponential bosonic operators, so we provide more details here.
To utilize \cref{eq:coherentidentity}, we write the exponential term in the kinetic energy operator as
\begin{align}\nonumber
&e^{\frac{1}{\sqrt{2}}(t_j(\hat{b}_j^\dagger-\hat{b}_j) - t_i (\hat{b}_i^\dagger - \hat{b}_i))}\\ %\nonumber
%&=
%e^{\frac{1}{\sqrt{2}}t_j(\hat{b}_j^\dagger-\hat{b}_j)}
%e^{-\frac{1}{\sqrt{2}}t_i (\hat{b}_i^\dagger - \hat{b}_i)}\\
&=
e^{\frac{1}{\sqrt{2}}(t_j\hat{b}_j^\dagger-t_i\hat{b}_i^\dagger)}
e^{\frac{1}{\sqrt{2}}(t_i \hat{b}_i-  t_j\hat{b}_j)}
e^{-\frac{1}{4}(t_i^2+t_j^2)},
\label{eq:lfkin}
\end{align}
where we have used
\begin{equation}
e^{\frac{1}{\sqrt{2}}t_j(\hat{b}_j^\dagger-\hat{b}_j)}
=e^{\frac{1}{\sqrt{2}}t_j\hat{b}_j^\dagger}
e^{-\frac{1}{\sqrt{2}}t_j\hat{b}_j}
e^{-\frac{1}{4} t_j^2} .
\end{equation}
The expectation value of \cref{eq:lfkin} is simple to evaluate with the semiclassical state of \cref{eq:cs} 
\begin{align}\nonumber
&\langle \Psi_T | e^{\frac{1}{\sqrt{2}}(t_j\hat{b}_j^\dagger-t_i\hat{b}_i^\dagger)}
e^{\frac{1}{\sqrt{2}}(t_i \hat{b}_i-  t_j\hat{b}_j)}| \Psi_T \rangle
e^{-\frac{1}{4}(t_i^2+t_j^2)}\\
&=
e^{\frac{1}{\sqrt{2}}(t_j\beta_j-t_i\beta_i)}
e^{\frac{1}{\sqrt{2}}(t_i \beta_i-  t_j\beta_j) }%(t_i \hat{b}_i-  t_j\hat{b}_j)}| \Psi_T \rangle
e^{-\frac{1}{4}(t_i^2+t_j^2)} .
\end{align}
}

The variational LF wavefunction
is expected to be more accurate than the semiclassical state due to the %presence of the 
explicit correlation between electrons and phonons.
Furthermore, the limit of $\omega \rightarrow \infty$ which is difficult for simple semiclassical wavefunctions to treat, can be exactly treated by the LF wavefunction, because the el-ph coupling term
in
$\hat{H}^{\text{LF}}$ can be removed by setting $\xi_i = \sqrt{2}g/\omega$.
Due to the fact that phonon displacements are significantly penalized in this limit, the variational optimization over $\pmb \beta$ naturally yields $\pmb \beta = \pmb 0$. 
Therefore, the bosonic operators in the hopping amplitude in \cref{eq:lfel} all vanish.
Provided that one can handle the remaining electronic Hamiltonian terms exactly, the variational LF wavefunction should be exact in this limit. 
We note that for many-electron systems in the $\omega \rightarrow \infty$ limit, the LF Hamiltonian takes the same form as the
attractive Hubbard model, which is another sign-free lattice model that can be efficiently simulated in AFQMC.\cite{shi2016infinite,shi2017many}

Despite these desirable properties, there seems to be no simple and general way to use this wavefunction in AFQMC without invoking 
a major increase in scaling.
%an exponential scaling either in the local energy evaluation or in the importance sampling. 
As an exception to this, we mention here the work of Hohenadler and co-workers\cite{Hohenadler2004} where a QMC algorithm with the LF Hamiltonian was formulated for single electron problems.
%Here i
It was demonstrated, however, that the transformed electronic Hamiltonian in \cref{eq:lfel} creates a complex phase problem.

Therefore, we briefly investigate a simpler linearized LF (LLF) wavefunction of the form,
\begin{equation}
|\Psi_\text{LLF}\rangle
=
(1 + \frac1{\sqrt{2}}\sum_i\xi_i \hat{n}_i \hat{b}_i^\dagger)
|\psi_T(\pmb\theta)\rangle
\otimes
|\phi_T(\pmb\beta)\rangle,
\label{eq:llfstate}
\end{equation}
where
we have omitted a term that is proportional to $\hat{n}_i\hat{b}_i$ since the action of $\hat{b}_i$ on $|\phi_T(\pmb\beta)\rangle$ is trivial due to \cref{eq:coherentidentity}.
We variationally optimize $\pmb \xi$ in \cref{eq:llfstate} to maximize the accuracy of the LLF trial wavefunction.
The AFQMC algorithm presented in \cref{sec:cpmc} can be efficiently implemented for \cref{eq:llfstate}.

It is possible to formulate a simple extension of the LLF wavefunction in the spirit of the TP wavefunction:
\begin{equation}
|\Psi_\text{TP-LLF}\rangle
=
\sum_{\alpha=1}^n c_\alpha
|\Psi_\text{LLF}^{(\alpha)}\rangle ,
%(1 + \frac1{\sqrt{2}}\sum_i\xi_i^{(\alpha)} \hat{n}_i \hat{b}_i^\dagger)
%|\psi_T(\pmb\theta^{(\alpha)})\rangle
%\otimes
%|\phi_T(\pmb\beta^{(\alpha)})\rangle,
\end{equation}
where each of the $|\Psi_\text{LLF}^{(\alpha)}\rangle$ terms has its own variational parameters.
Following the discussion of the TP trial wavefunction, it may be possible to determine these variational parameters via
a convex interpolation of $\pmb \beta$ and $\pmb \xi$ as in \cref{eq:interp}.
We refer to this wavefunction as the TP-LLF($n$) wavefunction which goes beyond both the TP$(n)$ and the LLF wavefunctions in sophistication.

In contrast with the LF form, a trial wave function with an el-ph Jastrow factor can be used more straightforwardly in AFQMC, since
$\hat{X}_i$ operators are involved in the exponent instead of 
$\hat{P}_i$ as in LF.
The unitary transformation in the LF wave function can be thought of as a simple Jastrow factor that
encodes correlation between the electronic density and the phonon momentum on a site.
However, 
%we note that unlike Jastrow factors (which are not unitary)
$\hat{U}_\text{LF}(\pmb \xi)$ is unitary and we thus expect this transformation to behave differently from %compared to the use of 
Jastrow factors in el-ph problems.\cite{ohgoe2014variational,ohgoe2017competition,karakuzu2017superconductivity} 
(It is also different from the coupled-cluster operators considered in recent studies of el-ph problems .\cite{sibaev2020molecular,dresselhaus2020coupling,white2020coupled})
Given the performance improvement with the LLF trial wave function (as discussed below), we 
expect an el-ph Jastrow trial wave function will 
greatly reduce the difficulties in 
 parameter regimes with strong el-ph coupling, and
result in a major improvement in our AFQMC approach. 
We leave the implementation and systematic studies using an el-ph Jastrow trial wavefunction in AFQMC for future work.  

\subsection{Additional details}

 The semiclassical state in \cref{eq:cs} can describe
two competing phases, SDW and CDW.  
To obtain the variational wavefunction for these two distinct states, we employ the following protocol:
\begin{enumerate}
\item For a CDW state, we perform a variational optimization of a semiclassical state with spin-restriction. Due to the spin-restriction, any states that arise from minimization are not capable of describing SDW order. 
\item For an SDW state, we perform a variational optimization of a spin-unrestricted Hartree-Fock (UHF)
wavefunction to minimize the electronic energy.  Once a UHF state is obtained, we determine the shift vector $\pmb \beta$ variationally while fixing the electronic degrees of freedom. As long as the UHF state exhibits SDW order, such a coupled el-ph semiclassical state with exhibit the same SDW order.
We have used an {\em ad hoc} effective repulsion strength ($U_\text{eff}/t$) of 0.5 \cite{Shi2013} in our UHF calculations
to obtain SDW trial wavefunctions for the Hubbard-Holstein model in this work. The CP-AFQMC results are not sensitive to this particular choice.
(We note that it is possible to determine this effective repulsion strength via a self-consistent procedure with CP-AFQMC \cite{qin2016coupling}.)
\end{enumerate}

\section{Perturbation Theory}
\subsection{Coherent State M{\o}ller-Plesset Perturbation Theory}\label{sec:csmp2}
It is instructive to %assess the performance of CP-AFQMC against 
consider low-order perturbation theory for comparison to numerically exact approaches.
We first note that
\begin{align}
\hat{\mathcal D}(\vec{\beta})^\dagger b_i^\dagger \hat{\mathcal D}(\vec{\beta}) &= b_i^\dagger + \beta_i,\\
\hat{\mathcal D}(\vec{\beta})^\dagger b_i \hat{\mathcal D}(\vec{\beta}) &= b_i + \beta_i.
\end{align}
Using this property, we write
\begin{align}
\hat{\bar{\mathcal H}}_\text{ph} \equiv
\hat{\mathcal D}(\vec{\beta})^\dagger \hat{\mathcal H}_\text{ph} \hat{\mathcal D}(\vec{\beta})
&=\omega \sum_i (b_i^\dagger + \beta_i) (b_i + \beta_i), \\
\end{align}
and
\begin{align}
\hat{\bar{\mathcal H}}_\text{el-ph} \equiv
\hat{\mathcal D}(\vec{\beta})^\dagger \hat{\mathcal H}_\text{el-ph} \hat{\mathcal D}(\vec{\beta})
&=
-g \sum_i
\hat{n}_i
(\hat{b}_i + \hat{b}_i^\dagger + 2 \beta_i).
\end{align}
Thus we have
\begin{align}
\hat{\bar{\mathcal H}}
=
\hat{\mathcal H}_\text{el}
+
\hat{\bar{\mathcal H}}_\text{ph}
+ \hat{\bar{\mathcal H}}_\text{el-ph}.
\end{align}
We note that the following zeroth-order Hamiltonian naturally has the semiclassical state of \cref{eq:cs} as its ground state
\begin{equation}
\hat{\mathcal H}_0
= \hat{\mathcal F} + \omega \sum_i (\beta_i^2  + \hat{b}_i^\dagger \hat{b}_i),% -2g \sum_i \hat{n}_i \beta_i
\end{equation}
where $\hat{\mathcal F}$ is the Fock operator defined as (for spin $\sigma=\uparrow$ or $\downarrow$)
\begin{align}
\hat{\mathcal F}_\sigma &= 
\hat{\mathcal F}_\sigma^\text{el}
-2g \sum_i \hat{n}_{i_\sigma} \beta_i,
\end{align}
%where 
with the electronic Fock operators: % is (for each spin $\alpha$ and $\beta$),
\begin{align}
\hat{\mathcal F}_\uparrow^\text{el} 
&=
-t
\sum_{\langle ij\rangle}
\hat{a}_{i_\uparrow}^\dagger
\hat{a}_{j_\uparrow}
+ U 
\sum_i
\hat{n}_{i_\uparrow}
\langle\hat{n}_{i_\downarrow}\rangle_{\psi_T},\\
\hat{\mathcal F}_\downarrow^\text{el}
&=
-t
\sum_{\langle ij\rangle}
\hat{a}_{i_\downarrow}^\dagger
\hat{a}_{j_\downarrow}
+ U 
\sum_i
\hat{n}_{i_\downarrow}
\langle\hat{n}_{i_\uparrow}\rangle_{\psi_T}.
\end{align}
where
\begin{equation}
\langle\hat{n}_{i_\sigma}\rangle_{\psi_T} = \frac{\langle\psi_T | \hat{n}_{i_\sigma}|\psi_T\rangle}{\langle\psi_T|\psi_T\rangle}
\end{equation}
It is straightforward to show that $|\Psi_T\rangle$ is an eigenstate of $\hat{\mathcal H}_0$.
From this starting point, one can develop an order-by-order perturbation theory to capture all of the 
correlation effects among electrons and between electrons and phonons built through
\begin{align}\nonumber
\hat{V} &= \hat{\bar{\mathcal H}} - \hat{\mathcal H}_0\\
&=
(\hat{\mathcal H}_\text{el} - \hat{\mathcal F}_\text{el})
%+\omega
%\sum_i (\beta_i \hat{b}_i^\dagger + \beta_i \hat{b}_i)
%-g
%\sum_i \hat{n}_i(\hat{b}_i + \hat{b}_i^\dagger)
+\sum_i (\omega\beta_i - g\hat{n}_i)(\hat{b}_i + \hat{b}_i^\dagger) .
\label{eq:perturb}
\end{align}
We note that such a partitioning of the Hamiltonian resembles the widely used M{\o}ller-Plesset (MP) perturbation theory in quantum chemistry.\cite{shavitt2009many}
We refer this perturbation theory to as ``coherent state M{\o}ller-Plesset perturbation theory'' (CSMP) since a coherent state (or a semiclassical state) is
an eigenstate of the zeroth-order Hamiltonian. This was also recently discussed in the work of White and co-workers in the context of coupled-cluster theory.\cite{white2020coupled}

Similar to MP, %with the first-order perturbation correction
CSMP recovers the energy of the semiclassical state with the first-order perturbation correction,
\begin{align}
E^{(0)} + E^{(1)}
= \langle  \hat{\mathcal H}_\text{el}  \rangle_{\psi_T}
+ \omega \sum_i \beta_i^2
- 2g \sum_i \langle \hat{n}_i \rangle_{\psi_T} \beta_i,
\end{align}
where we have defined
\begin{equation}
\langle \hat{n}_i \rangle_{\psi_T} 
\equiv
\langle \psi_T | \hat{n}_{i_\uparrow} + \hat{n}_{i_\downarrow} | \psi_T \rangle.
\end{equation}
In this work, we are interested in comparing the second-order perturbation theory (CSMP2) with AFQMC. %CP-AFQMC.
The evaluation of the CSMP2 energy is most natural in the molecular orbital (MO) basis rather than in the site basis.
The MO basis is defined by a set of orbitals, $\{\psi_{i_\sigma}\}$, that satisfy
\begin{equation}
\hat{\mathcal F}_\sigma \psi_{p_\sigma}
=
\epsilon_{p_\sigma} \psi_{p_\sigma},
\end{equation}
where $\epsilon_{p_\sigma}$ is the $p$-th MO energy and the $p$-th MO,
$\psi_{p_\sigma}$, is expanded via a set of site orbitals, $\{\phi_{\mu_\sigma}\}$,
\begin{equation}
\psi_{p_\sigma}
=
\sum_\mu C_{\mu p_{\sigma}} \phi_{\mu_\sigma}.
\end{equation}
We then transform \cref{eq:perturb} from the site basis to the MO basis using the coefficient matrix $\mathbf C$ for each spin
\begin{align} \nonumber
\hat{\bar{V}}
&=
\sum_{\sigma\in\{\uparrow,\downarrow\}} \sum_{pq} 
\sum_\mu
\left(
( \omega \beta_\mu -g_{pq} )
\hat{a}_{p_\sigma}^\dagger \hat{a}_{q_\sigma} \hat{b}_\mu^\dagger + \text{h.c.}\right) \\
&+
\sum_{pqrs}
U_{p_\uparrow q_\downarrow r_\downarrow s_\uparrow} 
\hat{a}_{p_\uparrow}^\dagger
\hat{a}_{q_\downarrow}^\dagger
\hat{a}_{r_\downarrow}
\hat{a}_{s_\uparrow},
\end{align}
where
\begin{align}
%\beta_r &= \sum_\mu (C_{\mu r})^* \beta_\mu \\
g_{pq}
&=
g \sum_\mu
(C_{\mu p})^*
C_{\mu q},
%(C_{\mu r})^*
\\
U_{p_\uparrow q_\downarrow r_\downarrow s_\uparrow} 
&=
U
\sum_{\mu}
(C_{\mu p_\uparrow})^*
(C_{\mu q_\downarrow})^*
C_{\mu r_\downarrow}
C_{\mu s_\uparrow}.
\end{align}

The CSMP2 energy expression follows in a spin-orbital MO basis,
\begin{align}\nonumber
E^{(2)}
&=
-\sum_{\mu=1}^{M}
\frac{(\omega \beta_\mu - g \sum_i (C_{\mu i})^* C_{\mu i}
)^2}{\omega} \\ \nonumber
&
- 
%\sum_{\mu=1}^{M}
\sum_{ia}
\frac{-|
%g 
%(C_{\mu a})^*
%C_{\mu i}
g_{ai}
|^2}
{\omega + \epsilon_a - \epsilon_i}\\
&-
\sum_{i_\uparrow a_\uparrow }
\sum_{j_\downarrow b_\downarrow}
\frac{
|
U_{i_\uparrow j_\downarrow b_\downarrow a_\uparrow}
|^2}
{\epsilon_{a_\uparrow} + \epsilon_{b_\downarrow} - \epsilon_{i_\uparrow} - \epsilon_{j_\downarrow}},
%\frac
%{
%|
%}
\label{eq:csmp2}
\end{align}
%\begin{align} \nonumber
%E^{(2)}
%&=
%-\sum_k \frac{(\omega\beta_k - g \langle \psi_T | \hat{n}_{k_\alpha} +\hat{n}_{k_\beta} | \psi_T \rangle)^2}{\omega}
%\\\nonumber
%&- g^2 \sum_{ia}
%\frac{|\sum_k \langle \psi_{i}^a | \hat{n}_{k_\alpha} + \hat{n}_{k_\beta} | \psi_T \rangle|^2}{\omega + \epsilon_a - \epsilon_i}\\
%&-\frac{U^2}4
%\sum_{ijab}
%\frac{
%|\sum_k\langle \psi_{ij}^{ab} | \hat{n}_{k_\alpha} \hat{n}_{k_\beta} | \psi_T\rangle|^2}
%{\epsilon_a + \epsilon_b - \epsilon_i - \epsilon_j}
%\end{align}
where the orbital energies $\{\epsilon_p\}$ are eigenvalues of the Fock operator $\hat{\mathcal F}$.
%This CSMP2 expression will help us gauge the accuracy of the CP-AFQMC results when other unbiased results obtained from
%approaches such as DMRG are not available.
We note that the first term in \cref{eq:csmp2} is zero if the semiclassical reference state is fully optimized.

\subsection{Lang-Firsov Perturbation Theory}\label{sec:lfpt2}
It may be useful to develop a second-order perturbation theory from a reference state given by the Lang-Firsov transformation\cite{lang1963kinetic} in \cref{eq:lfstate}.
%This was done first by Bonca for the 2-electron Hubbard-Holstein model in the continuum limit (infinite lattice limit).
%In particular, Bonca and co-workers chose
%We start from the fact that 
%\begin{align}
%\hat{U}(\pmb \xi)^\dagger
%a_{i_\sigma}^\dagger
%a_{j_\sigma}
%\hat{U}(\pmb \xi)
%&=
%a_{i_\sigma}^\dagger
%a_{j_\sigma}
%e^{(-\xi_i(b^\dagger_i-b_i)+ \xi_j(b^\dagger_j-b_j))},\\
%\hat{U}(\pmb{\xi})^\dagger b_i^\dagger \hat{U}(\pmb \xi) &= b_i^\dagger - \xi_i\hat{n}_i,\\
%\hat{U}(\pmb{\xi})^\dagger b_i \hat{U}(\pmb \xi) &= b_i - \xi_i\hat{n}_i.
%\end{align}
%The similarity transformed Hamiltonian reads
%\begin{align}\nonumber
%\hat{{\mathcal H}}_\text{el}^\text{LF}
%&=
%-t \sum_\sigma \sum_{\langle ij \rangle} 
%a_{i_\sigma}^\dagger
%a_{j_\sigma}
%e^{(-\xi_i(b^\dagger_i-b_i) +\xi_j(b^\dagger_j-b_j))}\\
%&+
%U\sum_i
%\hat{n}_{i_\uparrow}
%\hat{n}_{i_\downarrow},\\
%\hat{{\mathcal H}}_\text{ph}^\text{LF}
%&=
%\omega
%\sum_i(\hat{b}^\dagger_i + \xi_i\hat{n}_i)(\hat{b}_i + \xi_i\hat{n}_i), \\ 
%\hat{{\mathcal H}}_\text{el-ph}^\text{LF}
%&=
%-g
%\sum_i\hat{n}_i(\hat{b}_i + \hat{b}_i ^\dagger + 2\xi_i\hat{n}_i ).
%\end{align}
In the spirit of the original LF transformation,\cite{lang1963kinetic}
we set
\begin{equation}
\xi_i = \frac {\sqrt{2}g}{ \omega},
\end{equation}
which removes the Holstein coupling in the transformed framework.
With this choice of the LF amplitudes \insertnew{instead of the variational LF amplitudes}, the transformed Hamiltonian is simplified to
\begin{align}\nonumber
\hat{\mathcal H}^\text{LF}
&=
\omega \sum_i \hat{b}_i^\dagger \hat{b}_i
-\frac{g^2}\omega \sum_i \hat{n}_i+ (U - \frac{2g^2}\omega) \sum_i \hat{n}_{i_\uparrow}\hat{n}_{i_\downarrow}\\
&-t \sum_\sigma \sum_{\langle ij \rangle} 
a_{i_\sigma}^\dagger
a_{j_\sigma}
e^{(-\frac{g}{\omega}(b^\dagger_i-b_i) +\frac{g}{\omega}(b^\dagger_j-b_j))}.
\label{eq:lfham}
\end{align}
For convenience, 
we rewrite
\begin{equation}
e^{(-\frac{g}{\omega}(b^\dagger_i-b_i) +\frac{g}{\omega}(b^\dagger_j-b_j))}
=
e^{-\frac{g^2}{\omega^2}}
e^{-\frac{g}{\omega}(b^\dagger_i-b^\dagger_j)}
e^{\frac{g}{\omega}(b_i-b_j)}.
\end{equation}
Based on the approach of Bonca, Trugman, and co-workers,\cite{bonca2000mobile,bonvca2000mobilea}
we choose the zeroth order Hamiltonian as
\begin{equation}
\hat{\mathcal H}_0^\text{LF} = 
\omega \sum_i \hat{b}_i^\dagger \hat{b}_i
-\frac{g^2}\omega \sum_i \hat{n}_i+ (U - \frac{2g^2}\omega) \sum_i \hat{n}_{i_\uparrow}\hat{n}_{i_\downarrow},
\end{equation}
with the perturbing Hamiltonian 
\begin{equation}
\hat{V}^\text{LF}
=
-t e^{-\frac{g^2}{\omega^2}}
\sum_\sigma \sum_{\langle ij \rangle} 
a_{i_\sigma}^\dagger
a_{j_\sigma}
e^{-\frac{g}{\omega}(b^\dagger_i-b^\dagger_j)}
e^{\frac{g}{\omega}(b_i-b_j)}.
\end{equation}

For concreteness, we consider the specific case of second-order perturbation theory for a two-electron, two-site problem. 
The ground state of $\hat{\mathcal H}_0^\text{LF}$
is
either
$|\uparrow_1\downarrow_1\rangle \otimes | 0, 0\rangle$ or $|\uparrow_2\downarrow_2\rangle \otimes | 0, 0\rangle$
with an energy $U - \frac{4g^2}{\omega}$.
We start from an unperturbed state,
\begin{equation}
|\Psi^{(0)}_0\rangle = 
\frac{1}{\sqrt{2}} (|\uparrow_1\downarrow_1\rangle \otimes | 0, 0\rangle,
+
|\uparrow_2\downarrow_2\rangle \otimes | 0, 0\rangle),
\end{equation}
noting that an excited state which can be connected to the unperturbed ground state via $\hat{V}^\text{LF}$ takes
the form 
\begin{equation}
|\Psi_{mn}^{(0)}\rangle
=
\frac{1}{\sqrt{2}}(|\uparrow_1\downarrow_2\rangle - |\downarrow_1\uparrow_2\rangle)
\otimes |m,n\rangle.
\end{equation}
It is then easy to show that the first-order energy correction to the unperturbed state is zero.
The second-order energy reads
\begin{align}\nonumber
&E^{(2)}_\text{LFPT}
=
-2t^2 e^{-\frac{2g^2}{\omega^2}}\\
&\times 
\sum_{m=0}^{\infty}
\sum_{n=0}^{\infty}
\frac{(g/\omega)^{2(m+n)}}{m!n!}
\frac{1+(-1)^{m+n}
}{(m+n)\omega - U + 2 \frac{g^2}{\omega}}.
\end{align}
We numerically evaluate this expression in a brute-force manner, observing that a maximum $n$ of 200 is sufficient to converge the energy to machine precision.
We note that this expression differs from that of Bonca, Trugman and co-workers \cite{bonca2000mobile,bonvca2000mobilea} since in their work a single bipolaron was considered in the continuum limit (infinite lattice) whereas in our work we focus on a 2-site problem. 

\section{The Holstein Model}\label{sec:holstein}
To study the behavior of %understand the scope of 
the proposed AFQMC algorithm with simple trial wavefunctions 
such as the semiclassical, LLF, and TP wavefunctions, 
we shall investigate
the 1D and 2D Holstein models first, namely we set $U=0$ in \cref{eq:HH}.

 For the Holstein model, the  sign problem is absent, as is 
well-known in the determinant quantum Monte Carlo (DQMC) approach \cite{batrouni2019langevin}.
Similar to DQMC, the overlap function 
$\langle \Psi_T | \psi_i(\tau), \mathbf X_i(\tau)\rangle$ in  \cref{eq:walkers} remains non-negative throughout the imaginary-time propagation, since
the  phonon component, $\phi_T(\mathbf X(\tau))$ (omitting the walker index again), is positive everywhere, and the electronic component, 
$\langle \psi_T | \psi \rangle =  | \langle \psi_{T,\uparrow} | \psi_\uparrow \rangle|^2$ with a spin-restricted form, is also non-negative. 
Thus, in the Holstein model the difference between our approach and DQMC is primarily in the way the Monte Carlo sampling is conducted. 
AFQMC uses a branching random walk with a population of walkers to construct the imaginary-time path iteratively, as we have described, while 
DQMC treats the entire path as a path integral or worldline, and updates it by sweeping different imaginary-time locations via a Metropolis-like 
algorithm. 
%\joonho{Shiwei, what do you mean by ``similarity transformation'' here?} 
A second difference is the introduction of an importance functions in our approach via the similarity transformation, 
as indicated in Eqs.~(\ref{eq:ph-dmc}) and (\ref{eq:oratio}). 
These factors can affect 
the behavior of the Monte Carlo sampling, and yield different performances in different regimes of the parameter space, including efficiency, 
auto-correlation time, and possibly different levels of difficulty with ergodicity. The examples in the Holstein model below serve as a first 
test of the AFQMC method in this context.

\subsection{2-Electron 2-site model}\label{ssec:2e2o}
We start %by assessing the accuracy of CP-AFQMC, CSMP2, and LFPT2 on a 
with this small problem where we easily can compare results
against exact diagonalization (ED).  Since there are only two sites in our model, we compute energies with open boundary conditions (OBCs). 
\begin{figure*}
\includegraphics[scale=0.75]{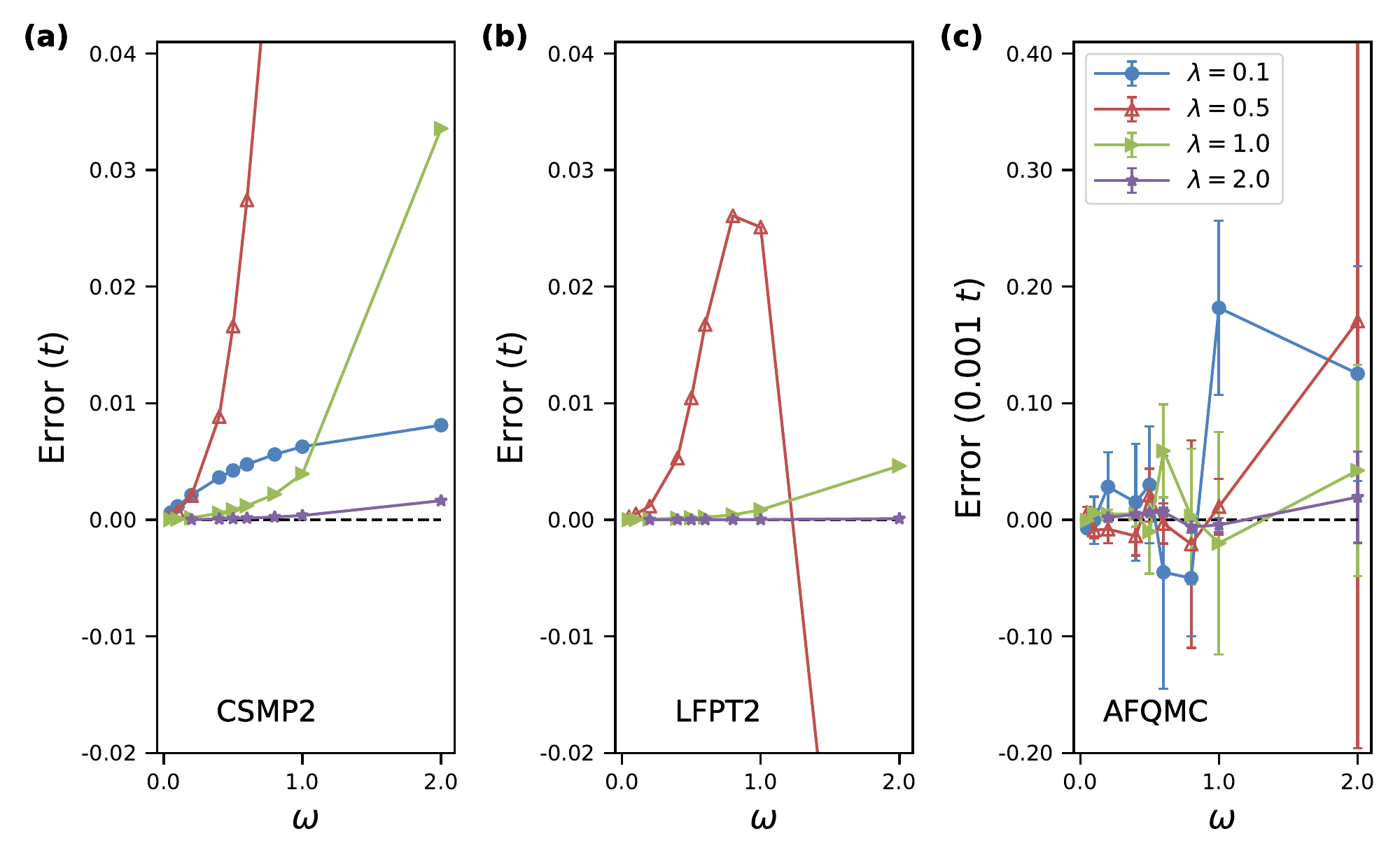}
\caption{\label{fig:2site}
Error in the total energy in units of $t$ for the 2-site 2-electron Holstein model as a function of $\omega$ for various $\lambda$ values:
(a) CSMP2, (b) LFPT2 and (c) AFQMC results. %Note that f
For $\lambda = 0.1$, LFPT2 energy errors lie outside the plotted range. 
In (c), for $\lambda\ge0.5$ AFQMC/TP(11) results are shown while for $\lambda = 0.1$ we present AFQMC/S results.
Note the different vertical scales in panel (c).
%\REMARKS{Sugges we combine with FIG 3 - see below}
}
\end{figure*}
In \cref{fig:2site}, we present
the error in the total energy of CSMP2, LFPT2 and AFQMC
compared to ED.
Understanding the behaviors of the two flavors of  perturbation theory helps gauge non-perturbative effects in our system.
% of CSMP2 is valuable in estimating the scope of CP-AFQMC with a single semiclassical trial wavefunction.
%In addition, LFPT2 is a standard perturbation theory, which makes it a useful benchmark to assess the ability of CP-AFQMC to capture non-perturbative effects.
In \cref{fig:2site}(a), it is clear that the CSMP2 energy becomes more inaccurate as we increase $\omega$. 
This is because the zeroth order wavefunction, a semiclassical trial wavefunction, starts to degrade when increasing el-ph correlation. 
Perhaps the most striking behavior to note concerning CSMP2 is that this approach performs worst for intermediate $\lambda$ values, (e.g. $\lambda = 0.5$) and is in fact more accurate for larger $\lambda$ values such as $\lambda = 2.0$. This can also be understood in terms of the increase in el-ph correlation as explained in \cref{ssec:semi}.
On the other hand, LFPT2 in \cref{fig:2site}(b) is comparatively more accurate than CSMP2 for $\lambda$ values larger than $\lambda = 0.5$. The LF reference state (namely two electrons occupying one site and with a phonon vacuum state) is qualitatively incorrect when the el-ph coupling is small. In such cases, we cannot treat the the hopping term perturbatively.
This is clearly reflected in \cref{fig:2site}(b) as LFPT2 exhibits large errors for small $\lambda$ values.
As LFPT2 is well known to produce accurate results for strong coupling, it is remarkable that a weak coupling perturbation theory, CSMP2, performs equally well even at $\lambda = 2.0$.

%\begin{figure*}
%%\includegraphics[scale=0.75]{/Users/joonholee/Dropbox/papers/inprep/holstein/2site/plots/lambda_cpmc.pdf}
%\includegraphics[scale=0.75]{plot/lambda_cpmc.pdf}
%\caption{\label{fig:2sitecpmc}
%Error in the total energy in units of $t$ and 0.001 $t$ for the 2-site 2-electron Holstein model as a function of $\omega$ for various $\lambda$ values.
%(a) CP-AFQMC results and (b) CP-AFQMC results (solid) and CP-AFQMC(11) results (dotted) are plotted. 
%In (b), for $\lambda < 0.5$ only CP-AFQMC results are presented and for $\lambda\ge0.5$ only CP-AFQMC(11) results are shown.
%\REMARKS{Suggest we remove panel a, label panel b by AFQMC and combine it with FIG 2 to make a 3-panel plot.}
%}
%\end{figure*}

%In \cref{fig:2sitecpmc}, we show 
We also show the performance of AFQMC  for the same 2-site Holstein dimer. % with a single semiclassical trial wave function as well as with TP($11$). %CP-AFQMC and CP-AFQMC(11) for the same 2-site Holstein dimer. 
The error of AFQMC %CP-AFQMC and CP-AFQMC(11) 
is shown on a much smaller (100 times) scale. % than that of \cref{fig:2site}.
\COMMENTED{
Comparing \cref{fig:2site}(a) and \cref{fig:2sitecpmc}(a), we recognize a similar behavior of the error between CSMP2 and CP-AFQMC at $\lambda = 0.5$. In particular, both exhibit larger errors for larger $\omega$ values. However, the magnitude of the error is far smaller for CP-AFQMC, with CP-AFQMC exhibiting near-exact energies for $\lambda < 0.5$, unlike CSMP2.
\cref{fig:2sitecpmc}(b) shows CP-AFQMC and CP-AFQMC(11) errors 
on a much smaller ($\sim$ 50 times) scale. % than that of \cref{fig:2site}.
}
We have tested both the single semiclassical state and TP($11$), i.e., the TP state with a superposition of 11 semiclassical states, as  trial wave function. 
These are referred to as AFQMC/S and AFQMC/TP($11$), respectively. Results are shown in the figure, with AFQMC/S for smaller $\lambda$ and 
 AFQMC/TP($11$)  for $\lambda \ge0.5$.  Near-exact energies are obtained for all parameters examined here. 
We observed that results can become severely biased  with AFQMC/S for  large $\lambda$, as a consequence of a poor importance function 
causing  large, or even diverging, variances. Even with an improved importance function TP($11$), small residual effects can be present (via underestimation of the statistical error, or bias from population size). \insertnew{We also note a large statistical error at $\omega = 2t$ which is maximized at an intermediate value of $\lambda = 0.5$ (or $g=\sqrt{2}t$). Nevertheless, with TP(11), the bias (if any) is smaller than 0.001 $t$ for the Holstein dimer, which highlights the accuracy and sampling efficiency of AFQMC/TP(11).}
We discuss the issue of bias in AFQMC in sign problem-free models further in \cref{ssec:4e4o} and \cref{ssec:Discuss-AC-sampling}.
%
%\REMARKS{The red line, $\lambda=0.1$ looks the most off, e.g. at $\omega\in [0.4,1]$. This is probably a fixable residual bias.
%Can we pick a point and check population size extrapolation? Other run parameters?}

\subsection{1D 4-site model at half-filling}\label{ssec:4e4o}

\insertnew{
To further investigate the effect of the importance function on 
the sampling result and any potential bias, we consider 
a 1D 4-site Holstein model employing periodic boundary conditions (PBCs), 
 at half-filling with $\lambda = 0.5$, $\omega = 4t$, and $g = 2t$.
 We compute the ground state energies with AFQMC using 
 the following trial wave functions:
 a single semiclassical state (S); 
the TP wavefunction with 13 semiclassical states (TP(13)); the LLF wavefunction; and the TP-LLF wavefunction with two LLF states (TP-LLF(2)).
With AFQMC/TP-LLF(2), a ground-state energy of  $-10.293(2)$ is obtained, compared to the  exact result of $-10.292$ 
(obtained from DMRG using iTensor\cite{fishman2020itensor},
although ED can also be done here).
In contrast, a biased result is seen with each of the other forms of the trial wave function. The bias is about $0.4$\% relative to the exact 
result, using the computational parameters specified in Sec.~\ref{sec:comp-details}, and is essentially independent of  whether 
bifurcation is accounted for or not in the trial wave function.
These results suggest that, to remove the sampling bias in this parameter regime,
it is critical to capture in the importance function %deal with 
both  a means to overcome the adiabatic potential
bifurcation issue and treat el-ph correlation.
}

\subsection{1D 20-site model at half-filling}
Next, we discuss a  20-site 1D Holstein model at half-filling employing PBCs. %, namely a 20-site model.
ED is no longer feasible for systems of this size so we used the iTensor\cite{fishman2020itensor} package to perform DMRG calculations\cite{jeckelmann1998density}.
The DMRG calculations were carried out by placing alternating fermionic and bosnic sites on a 1D % one dimensional 
lattice so that overall there are twice the number of sites compared to the physical lattice problem. While it is possible to use an optimized phonon basis\cite{zhang1998density} to handle larger el-ph coupling cases, here we employed the most primitive version of DMRG for simple comparisons.  The bond dimension we used was fixed at 1000 and the maximum number of bosons for each site was taken to be 60. 
\begin{figure*}
\includegraphics[scale=0.75]{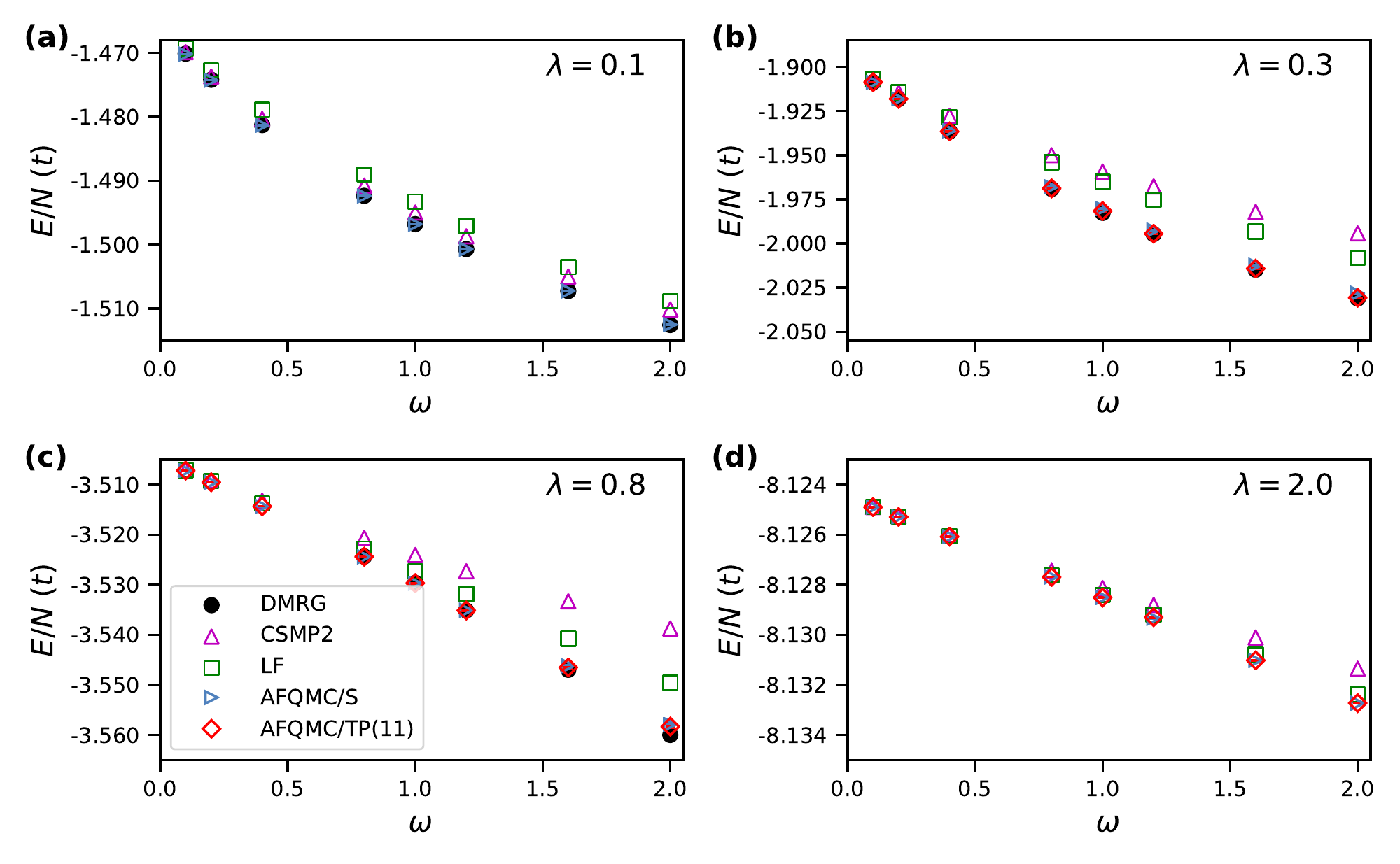}
\caption{\label{fig:20site}
Total energy per site in units of $t$ for the 20-site 20-electron 1D Holstein model as a function of $\omega$ for various values of $\lambda$ values: (a) $\lambda = 0.1$ results, (b) $\lambda = 0.3$ results, (c) $\lambda = 0.8$ results, and (d) $\lambda = 2.0$ results.
Note that all error bars for AFQMC are too small to be seen on the plotted scales,
DMRG reference values are unavailable for $\omega < 0.8$ when $\lambda = 0.8$ and $\lambda = 2.0$,
and AFQMC/TP(11) is not presented for $\lambda = 0.1$ because the results for this trial wavefunction are nearly identical to those of AFQMC/S.
}
\end{figure*}

We compare the total energy per site within DMRG, CSMP2, \insertnew{variational LF,} AFQMC/S, and AFQMC/TP($11$) %CP-AFQMC, and CP-AFQMC(11) 
in \cref{fig:20site} for various $\lambda$ and $\omega$. \insertnew{Given the discussion of \cref{ssec:4e4o}, it is desirable to employ the TP-LLF($n$) wavefunctions in general, but we leave a more detailed study with this trial wavefunction for a future study. Here, we focus on AFQMC with simpler and less accurate trial wavefunctions (AFQMC/S and AFQMC/TP(11)). } 

Similarly to the two-site problem, we observe that CSMP2 follows the (near-exact) answers given by DMRG and AFQMC %CP-AFQMC 
closely at small $\lambda$ (e.g. $\lambda = 0.1$) as the frequency of the phonon mode is varied.  However, we see a clear quantitative deviation of CSMP2 from the other curves as $\omega$ increases. The deviation is again maximized at an intermediate coupling $\lambda = 0.8$ and is smaller at weak and strong couplings. 
\insertnew{The variational LF wavefunction works better than CSMP2 for all $\lambda > 0.1$. Its strength over CSMP2 is highlighted as $\omega$ increases.
This clearly suggests that the variational LF wavefunction includes el-ph correlation beyond the second-order contribution provided in CSMP2.
}

%\REMARKS{adjust discussion below after final confirmation of results}
The performance of AFQMC/S is very good at all coupling strengths considered here for $\omega \le 2.0$. Similar issues with biased final estimates from 
poor importance functions are seen at $\lambda = 0.3$ and $\lambda=0.8$. 
AFQMC/TP($11$) shows improvement over the simplest semiclassical importance function in the case of $\lambda = 0.3$.
However its improvement for $\lambda = 0.8$ as $\omega$ becomes larger is very small. For example, the residual bias is still visible at  $\omega = 2.0$. 
This points to the need to improve the importance function over the forms we have used. \insertnew{We expect that incorporating el-ph correlation directly, as in the LF-type wavefunctions, will ameliorate this sampling bias greatly, as observed in \cref{ssec:4e4o}.}
 For the rest of the paper, we focus on AFQMC with \insertnew{the simplest trial wavefunction}, namely a single semiclassical trial wavefunction, \insertnew{because the observed sampling bias is small enough that it does not affect the conclusions of this work}.

%The largest error per site for CP-AFQMC(11) (0.0017(2)) occurs at $\lambda = 0.8$ and $\omega = 2.0$. This suggests that the nature of this parameter regime is not easily described by a linear combination of semiclassical states. Therefore, for the rest of the paper, we focus on CP-AFQMC with a single semiclassical trial wavefunction.

\COMMENTED{
The performance of CP-AFQMC with a single semiclassical state is 
significantly better than CSMP2. In particular, this approach appears to go far beyond simple perturbation theory and approximates well the DMRG results at all couplings considered here for $\omega \le 2.0$. 
However, we note that similarly to CSMP2, the deviation of CP-AFQMC from DMRG increases with $\omega$ and is maximized at an intermediate $\lambda = 0.8$. 
%It should be noted that, unsurprisingly, CP-AFQMC is quantitatively more accurate than CSMP2.
We also compare results for a TP trial wavefunction with a total of 11 semiclassical states, namely CP-AFQMC(11), to DMRG in \cref{fig:20site}. 
While CP-AFQMC(11) clearly shows some improvement over the most simple semiclassical trial in the case of $\lambda = 0.3$, its improvement for $\lambda = 0.8$ as $\omega$ becomes larger is very small. The largest error per site for CP-AFQMC(11) (0.0017(2)) occurs at $\lambda = 0.8$ and $\omega = 2.0$. This suggests that the nature of this parameter regime is not easily described by a linear combination of semiclassical states. Therefore, for the rest of the paper, we focus on CP-AFQMC with a single semiclassical trial wavefunction.
}

\subsection{2D 4x4 model at half-filling}
We have established the expected behavior of AFQMC with semicalssical wavefunctions as importance functions from studying one-dimensional problems such as the Holstein dimer and a 1D chain. 
Here, we explore higher dimensions by investigating a 2D square lattice problem with a 4x4 geometry.
We employed PBCs along x-axis and OBCs along y-axis.
The main reason for choosing this boundary condition is to ease the convergence of the DMRG calculations. 
We were able to converge DMRG calculations only for $0.8 \ge \omega$ and $\lambda \le 0.5$ where we used a bond dimension of 2500 and a maximum number of bosons of 25. 

\begin{figure*}
\includegraphics[scale=0.75]{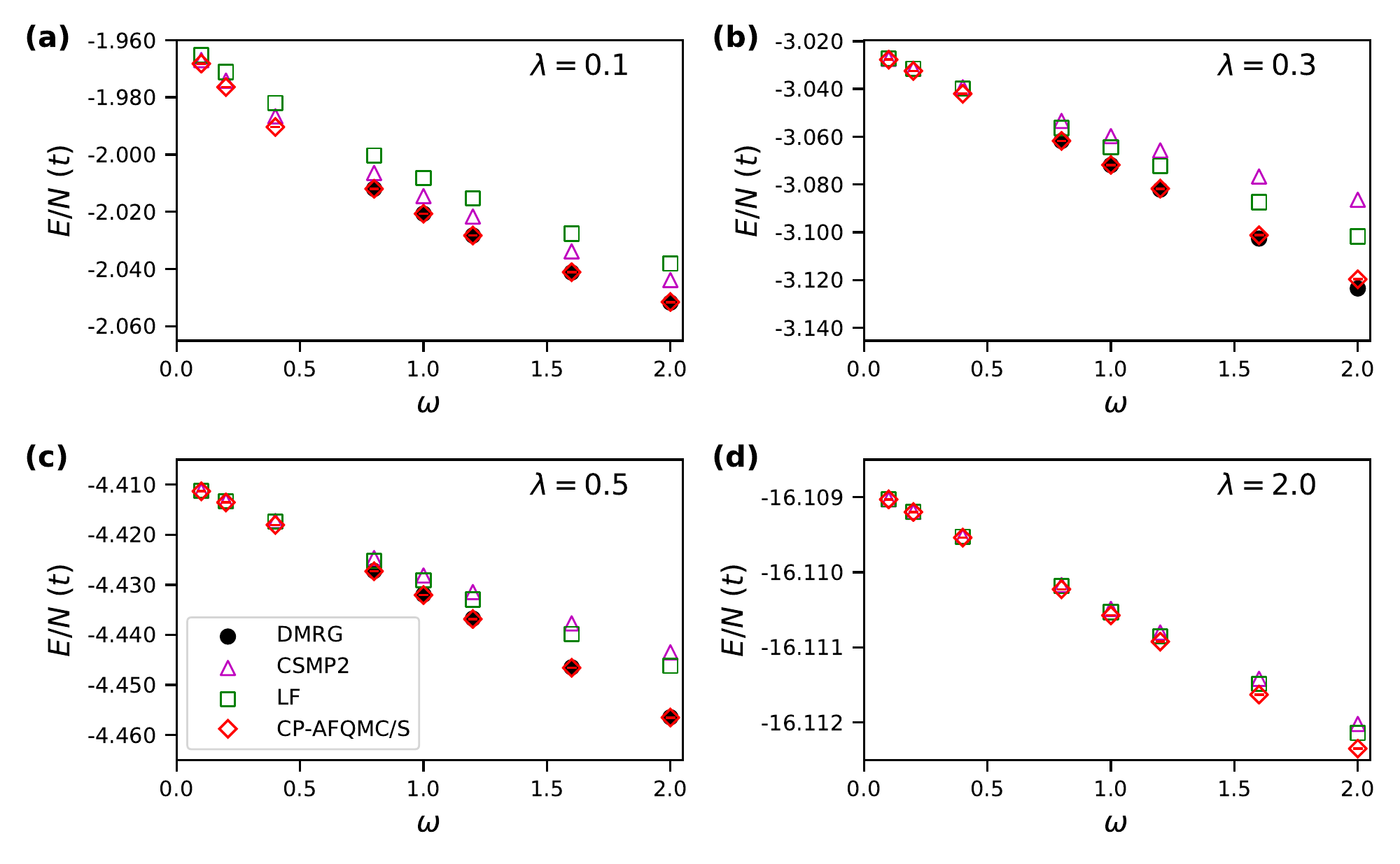}
\caption{\label{fig:4x4site}
Total energy per site in units of $t$ for the 4x4 2D Holstein model at half-filling as a function of $\omega$ for various $\lambda$ values: (a) $\lambda = 0.1$ results, (b) $\lambda = 0.3$ results, (c) $\lambda = 0.5$ results, and (d) $\lambda = 2.0$ results.
Note that all error bars of CP-AFQMC/S are too small to be seen on the plotted scales,
DMRG reference values are only available for $\omega \ge 0.8$ when $\lambda < 2.0$.
}
\end{figure*}

In \cref{fig:4x4site}, the energy per site as a function of $\omega$ for various $\lambda$ values is presented for this 2D model. 
We observe conclusions similar to our previous one-dimensional examples. CSMP2 quantitatively fails as $\omega$ increases. 
Furthermore, for a fixed $\omega$, CSMP2 performs worst for intermediate $\lambda$ values and is more accurate for small and large $\lambda$ values. 
\insertnew{Similarly to the 1D 20-site case, the variational LF energy is more accurate than CSMP2 for $\lambda > 0.1$ and its improvement over CSMP2 becomes larger as $\omega$ increases.}
AFQMC/S 
%generally follows this trend, but is much more quantitatively accurate than CSMP2. In fact, for the particular range $\omega \in [0.1, 2.0]$, it 
is well behaved in the range $\omega \in [0.1, 2.0]$. Its maximum error occurs at
 $\lambda = 0.3$ and $\omega = 2.0$,
 \insertnew{where clear indications of sampling bias arises.}
\insertnew{Nevertheless, the range of parameters where AFQMC/S can be reliably performed with the simplest possible semiclassical trial wavefunction is quite broad even in 2D, highlighting the utility and potential of this approach. }
 %where clear indication of the autocorrelation problem due to the poor importance function is seen.
% \REMARKS{pending update}
% 
% \COMMENTED{
% $\omega \in [0.1, 2.0]$
%nearly exact 
%at every $\lambda$ except $\lambda = 0.3$. The maximum deviation of energy per site of 0.00384(8) occurs at $\lambda = 0.3$ and $\omega = 2.0$. 
%The range of parameters where CP-AFQMC can be reliably performed with a simplest possible semiclassical trial wavefunction is quite broad even in 2D, highlighting the utility and potential of this approach. 
%}

%\subsection {Autocorrelation time}
\subsection {Autocorrelation time and variance control}
\label{ssec:Discuss-AC-sampling}

As we mentioned at the beginning of this section, in the Holstein model the difference between our AFQMC approach and DQMC is mainly in the details 
of the Monte Carlo sampling algorithm. The two methods can thus have different behavior in terms of efficiency in different regimes of the parameter space. 
Here we look into this to help understand the domain of applicability. We note that this is not the focus of our study, since in the most general case 
where electron interactions are present, the branching random walk approach must be adopted in order to control the 
sign or phase problem. 

The standard DQMC algorithm  based on local updates exhibits
a long autocorrelation time in the Holstein model for $\omega< 0.5$ and for low temperatures. % in general.
%This is quite problematic because the standard algorithm does not have a fermionic sign problem for the Holstein model.
%In principle, an exact simulation should be possible without the need for exponentially many walkers.
%However, it 
It has been found that this is a consequence of an ergodicity problem.
% arise for $\omega< 0.5$ and a long autocorrelation time is a manifestation of this.
A careful mathematical analysis of the causes of this problem can be found in the work of Hohenadler and co-workers.\cite{Hohenadler2004,Hohenadler2005,Hohenadler2007}
These authors have shown that the condition number of the bosonic action sampled in DQMC for small values of $\Delta\tau$
scales as $1/(\omega \Delta \tau)^2$. This poorly conditioned action leads to a long autocorrelation time that scales quadratically with \insertnew{increasing $\omega^{-1}$}.
%There have been quite a few approaches that attempt
There have been attempts to ameliorate this problem based on global moves such as the Langevin dynamics approach\cite{Karakuzu2018,batrouni2019langevin,Hebert2019} and the self-learning Monte Carlo approach.\cite{Chen2018} We also mention that the work of Hohenadler and co-workers %completely 
removed the autocorrelation problem using the Lang-Firsov transformation along with a principal component analysis.\cite{Hohenadler2004}

\begin{figure}
\includegraphics[scale=0.43]{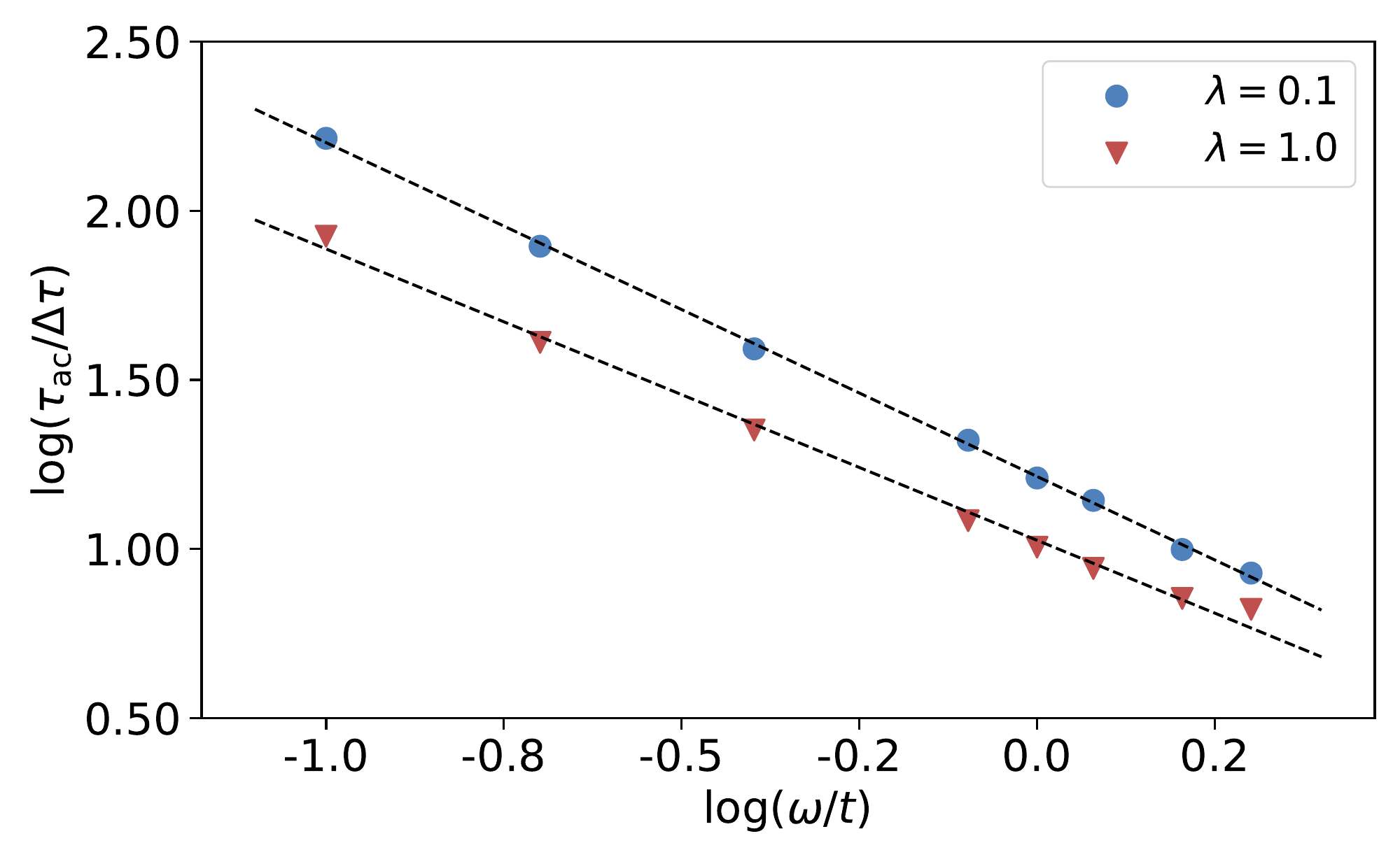}
\caption{\label{fig:autocorr}
Log-log plot of autocorrelation time $\tau_\text{ac} / \Delta\tau$ and phonon frequency $\omega/t$ for $\lambda = 0.1$ and $\lambda = 1.0$. The black dotted lines are linear fits for each curve.
For $\lambda = 0.1$, the slope is -0.9876 with $R^2 = 0.9993$ and for $\lambda = 1.0$, the slope is -0.8617 with $R^2 = 0.9942$. 
%Total energy per site in the unit of $t$ of the 20-site 20-electron 1D Holstein model as a function of $\omega$ for various $\lambda$ values: (a) $\lambda = 0.1$, (b) $\lambda = 0.3$, (c) $\lambda = 0.8$, and (d) $\lambda = 2.0$.
%Note that all error bars of CP-AFQMC are invisible in the plotted scales,
%DMRG reference values are unavailable for $\omega < 0.8$ when $\lambda = 0.8$ and $\lambda = 2.0$,
%and CP-AFQMC(11) is not presented for $\lambda = 0.1$ because the trial wavefunction is nearly identical to CP-AFQMC.
\COMMENTED{Can we scrutinize the $\lambda=1$ results and make sure the data are reliable? For any $\omega$ where we do not expect the 
result to be exact within error bars, the autocorrelation time is under-estimated. Again, this might not show up in a 'normal' run, if the semiclassical 
wf is biasing the sampling SO much} 
%\joonho{the sampling bias exists only for $\omega = 2.0$ and $\lambda = 1.0$ so we might consider removing that point, but I don't think that the fitted function would change much because the slope is largely determined by other data points.}
}
\end{figure}

%Since our approach is uniquely well suited for simulating systems with small values of $\omega/\lambda$, it is important to investigate
%this issue for small frequencies. 
Since AFQMC applies a projector to the entire set of electronic and phonon degrees of freedom using a population of random walkers, it is less prone to ergodicity problems. The Monte Carlo time coincides with  the imaginary-time direction, with open-ended random walks evolving along the worldines, which makes them less likely to become trapped in particular configurations of the phonon paths. 
There is a deep connection between this fact and the necessity to resort to this sampling approach in order to impose a CP or phaseless gauge condition \cite{Zhang1997,zhang2003quantum}. 
%namely global moves are employed at each time step, it is expected that a potential autocorrelation time problem would not be as severe for this algorithm.
%To numerically investigate this, 
To quantify this, we directly compute the energy autocorrelation function,
\begin{equation}
c_E(\tau) = \frac{1}{N-\tau}
\sum_{n=1}^{N-\tau}
(E_n - \langle E\rangle) (E_{n+\tau} - \langle E\rangle) ,
\end{equation}
which gives an estimate of the integrated autocorrelation time via
\begin{equation}
\tau_\text{ac}
=
\sum_{\tau=-\infty}^{\infty} \frac{c_E(\tau)}{c_E(0)} .
\label{eq:tau}
\end{equation}
The summation in \cref{eq:tau} needs to be performed within some window instead of over the entire set of samples since, for $\tau \gg \tau_\text{ac}$, the summed noise becomes comparable to the actual signal.  We follow Sokal's prescription of the automated windowing procedure to handle this issue.\cite{mackey,Sokal1997}
In \cref{fig:autocorr}, 
we present estimates of autocorrelation times for $\lambda = 0.1$ and $\lambda = 1.0$ for the 4x4 2D Holstein model.
For both values of $\lambda$, we observe a near linear behavior in the log-log scale correlation between $\tau_\text{ac}$ and $\omega$ as in \cref{fig:autocorr}. Empirically, we find that the autocorrelation time scales as $1/\omega^{0.9876}$ for $\lambda=0.1$ and $1/\omega^{0.8617}$ for $\lambda = 1.0$. 
This scaling is a significant improvement over that of the standard DQMC algorithms where $\tau_\text{ac}$ scales as $1/\omega^2$.\cite{Hohenadler2004}

On the other hand, in AFQMC we use an importance function to guide the random walks. If the quality of the importance function is very poor, the
variance can grow \insertnew{and even become infinite.\cite{shi2016infinite}} In cases where the trial wave function suppresses certain regions of the \insertnew{Hilbert} space being sampled with a qualitatively incorrect 
functional form, the autocorrelation time and thus the variance can diverge, as mentioned earlier in this section. This situation was seen in the examples with the semiclassical wave function
where there is a strong bifurcation of the adiabatic potential in the Holstein \insertnew{dimer}. 
%\insertnew{Another instance was that the increase in el-ph correlation makes the use of semiclassical trial wavefunctions and related ones introduce large sampling biases.}
Another example occurs with the use of semiclasscial trial wave functions where the lack of explicit el-ph correlation leads to large sampling biases.
In extreme cases, calculations will be seemingly well-behaved in ``normal-sized'' runs, as the Monte Carlo sampling is 
strongly biased by the wrong importance function and the auto-correlation time grows exponentially. 
These situations require careful analysis of the variance and study of the dependence on the details of the importance function to reveal the problem.\cite{shi2016infinite} 
%\joonho{Shiwei, could you elaborate more what you mean in the following sentences?}
\insertnew{Separate but related to the quality of the importance function is the issue of efficiently sampling of multi-modal landscapes
in the el-ph models, as we have only incorporated local moves in our
random walks. In the worst cases,  AFQMC 
can,
even with reasonable choices of importance functions,
experience  difficulties with long autocorrelations 
as occurs in DQMC.
In AFQMC the use of a population of open-ended random walkers  with branching can help avoid the sampling being stuck. 
}

\section{The Hubbard-Holstein Model}\label{sec:hh}

%The focus of our AFQMC method is on doped systems and more realistic Hamiltonians, where the sign problem or phase problem will be present.
%In the previous section, we studied the 1D and 2D Holstein models with simple trial wavefunctions. 
%In this section, we present benchmark data using the same approach for the Hubbard-Holstein model with $U/t=4$. 
%Because of the competition between $U$ and $g$, we carefully study trial wavefunctions with both CDW and SDW order.
%Note that the AFQMC %CP-AFQMC 
%algorithm is no longer exact even if the phonons are completely classical due to the fact that the electron-electron repulsion will lead to a sign problem.  Karakuzu, Seki, and Sorella have presented an efficient QMC algorithm which is free of the sign-problem at half-filling as long as $U > 2g^2/\omega$.\cite{Karakuzu2018}
%Within AFQMC, we can simulate any parameter regime without the sign problem at the expense of the constraint bias. 
%Since the significance of the AFQMC bias has been well-documented for electronic simulations in the past,\cite{Qin2016} we focus on any additional biases that arise from the interplay between electrons and phonons in this section.
%We use an ad hoc $U_\text{eff} = 0.5$ to generate all the trial wavefunctions for  AFQMC in this section.
The focus of our CP-AFQMC method is on doped systems and more realistic Hamiltonians, where the sign problem or phase problem will be present.
In the previous section, we studied the 1D and 2D Holstein models with simple trial wavefunctions. 
In this section, we present benchmark data using the same approach for the Hubbard-Holstein model with $U/t=4$. 
Because of the competition between $U$ and $g$, we carefully study trial wavefunctions with both CDW and SDW order.
Note that the CP-AFQMC %CPMC 
algorithm is no longer exact %even if the phonons are completely classical due to the fact that 
because the el-el repulsion will lead to a sign problem.  Karakuzu, Seki, and Sorella have presented an efficient QMC algorithm which is free of the sign-problem at half-filling as long as $U > 2g^2/\omega$.\cite{Karakuzu2018}
Within CP-AFQMC, we can simulate any parameter regime without the sign problem at the expense of the constraint bias. 
Since CP-AFQMC has been extensively benchmarked %the significance of the AFQMC bias has been well-documented 
for electronic systems in the past,\cite{Qin2016} we focus on any additional biases that may arise from the interplay between electrons and phonons in this section.
We use an {\em ad hoc} $U_\text{eff} = 0.5\,t$ in the electronic mean-field part to generate all the SDW trial wavefunctions for CP-AFQMC in this section.

Below, we first examine the behavior of our algorithm in 1D, and then in 2D both at half-filling and $1/8$-doping. We will focus on benchmarking the 
accuracy of the computed ground-state energies.  
We note that for the purely electronic cases with $\lambda=0$, all of our models in 1D and at half-filling in 2D are sign-problem-free. However, CP-AFQMC 
can incur a systematic error in the energy in these cases, because of an ``artificial node" in auxiliary-field space; \cite{Zhang1997} this error can be removed
straightforwardly. \cite{Shi2013,Qin2016} 
Instead of invoking the scheme to  remove this artificial bias, we will perform the CP-AFQMC calculation in the generic way as described above, since our main focus in this work is the most general 
situation of a doped Hubbard-Holstein model where the sign problem is present. 

\subsection{1D 20-site model at half-filling}
\begin{figure*}
\includegraphics[scale=0.75]{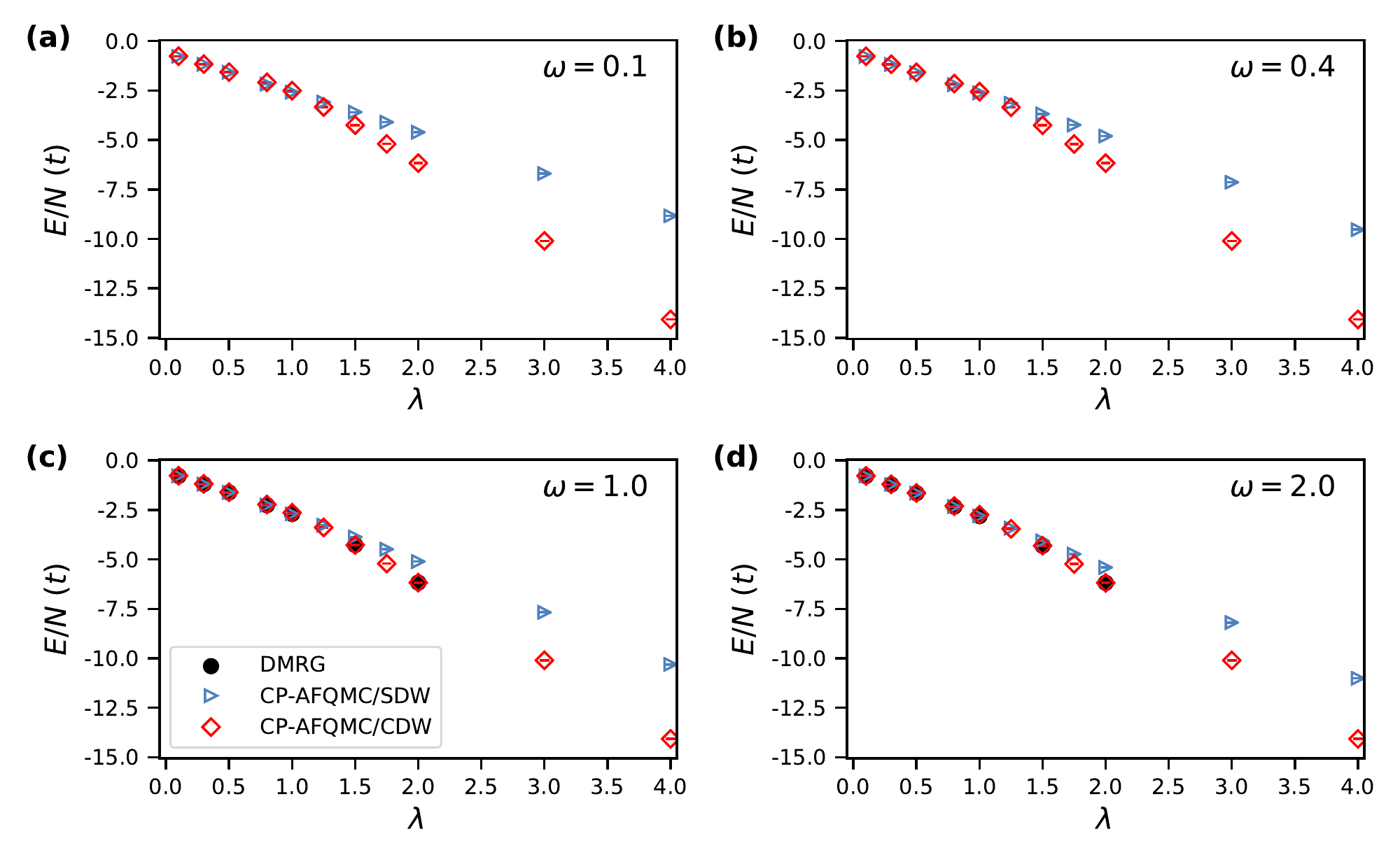}
\caption{\label{fig:20site_hh}
The total energy per site in units of $t$ for the 20-site 1D Hubbard-Holstein model at half-filling with $U=4t$ as a function of $\lambda$ for various $\omega$ values: (a) $\omega = 0.1$, (b) $\omega = 0.4$, (c) $\omega = 1.0$, and (d) $\omega = 2.0$.
%The solid and dotted lines are CP-AFQMC with SDW and CDW trial wavefunctions, respectively.
Note that all error bars of CP-AFQMC are too small to be seen on the plotted scales.
Note that the DMRG results are unavailable for $\omega = 0.1$ and $\omega = 0.4$ as well as for $\lambda > 2$. 
}
\end{figure*}

We benchmark CP-AFQMC against DMRG for the 20-site 1D Hubbard-Holstein model at half-filling with PBCs. 
Unlike for the case of the pure Holstein model, CSMP2 \insertnew{and the variational LF approach are} quantitatively and qualitatively inaccurate for all parameters examined here.
This is not surprising because the on-site repulsion term for $U/t=4$ is not small, so the failure of mean-field theories and a low-order perturbation theory on the el-el interaction is expected.  For this reason we do not discuss CSMP2 \insertnew{and variational LF} results here.

%Our simple mean-field trial wavefunction undergoes a phase transition between CDW and SDW phases at $\lambda = 0.5$ because $U_\text{eff}$ is 0.5. This makes the choice of the trial ambiguous for $\lambda \ge 0.5$ because we have two possible trial wavefunctions (CDW and SDW) that can be used in the CP-AFQMC calculations. We will consider CP-AFQMC with both trails and refer to them as CP-AFQMC(CDW) and CP-AFQMC(SDW), respectively.  Although CP-AFQMC calculations are not variational,\cite{carlson1999issues} we follow standard practice and choose the appropriate trial wavefunction for a given $\lambda$ based on relative energies.
To study the CDW and SDW phases and the possibility of a phase transition between them, we carry out CP-AFQMC calculations using two different mean-field 
trial wave functions with the corresponding broken symmetry. Comparison of the computed energies indicates which one is the ground state at  each Hamiltonian parameter choice as well as the existence and location of a transition,
 although the fact that our CP-AFQMC energies computed from the mixed estimate are not variational \cite{carlson1999issues} adds a subtlety to this procedure.
Here the calculation leading to the higher energy can be thought of as the constraint 
 acting to ``hold'' the projection to an excited state compatible with the broken symmetry of the trial wave function. In actual applications, we could 
 use a self-consistent CP-AFQMC procedure \cite{qin2016coupling} to tune the trial wave function and reduce its effect on the result, but   for the 
 purpose of benchmarks we will only perform one-shot calculations here using UHF trial wave functions generated with a fixed $U_{\rm eff}$, and rely on 
 comparison with DMRG results to gauge the accuracy.

%In \cref{fig:20site_hh}, for $\lambda \ge 0.5$ we compare two sets of CP-AFQMC results at different $\omega$ values; one set with SDW trial wavefunctions and another with CDW trial wavefunctions.
%For $\lambda \le 1.0$, we see that the CDW trials lead to higher CP-AFQMC energies than those of the SDW trials.
%The energy differences are large enough to make it straightforward to choose an appropriate trial wavefunction for a given set of parameters.
%\REMARKS{Check below. Original text seemed to indicate only one trial wf was used for   $\lambda < 0.5$ but two sets of symbols are shown in the figure}
%In \cref{fig:20site_hh}, for $\lambda \ge 0.5$ we compare two sets of CPMC results at different $\omega$ values; one set with SDW trial wavefunctions and another with CDW trial wavefunctions.
In \cref{fig:20site_hh}, we compare two sets of CP-AFQMC results at different $\omega$ values, one set with SDW trial wavefunctions (denoted by CP-AFQMC/SDW)  and another with CDW trial wavefunctions (denoted CP-AFQMC/CDW).
The trial wave functions themselves show a 
SDW to CDW transition 
at $\lambda \sim 0.5$, given our {\em ad hoc} choice of  $U_\text{eff}= 0.5$. 
We see that  CP-AFQMC/CDW leads to higher energies than CP-AFQMC/SDW for $\lambda \le 1.0$.
The energy differences are large enough to make it straightforward to identify the correct phase.

%\begin{figure*}
%\includegraphics[scale=0.75]{/Users/joonholee/Dropbox/papers/inprep/holstein/hubbard_holstein_20sites/plots/figure_20site_hh.pdf}
%\caption{\label{fig:20site_hh}
%Total energy per site in the unit of $t$ of the 20-site 1D Hubbard-Holstein model at half-filling with $u=4$ as a function of $\omega$ for various $\lambda$ values: (a) $\lambda = 0.1$, (b) $\lambda = 0.3$, (c) $\lambda = 0.8$, and (d) $\lambda = 2.0$.
%Note that all error bars of CP-AFQMC are invisible in the plotted scales, and
%the DMRG results are unavailable for $\lambda = 2.0$ at $\omega < 1.0$. 
%}
%\end{figure*}

\begin{figure*}
\includegraphics[scale=0.75]{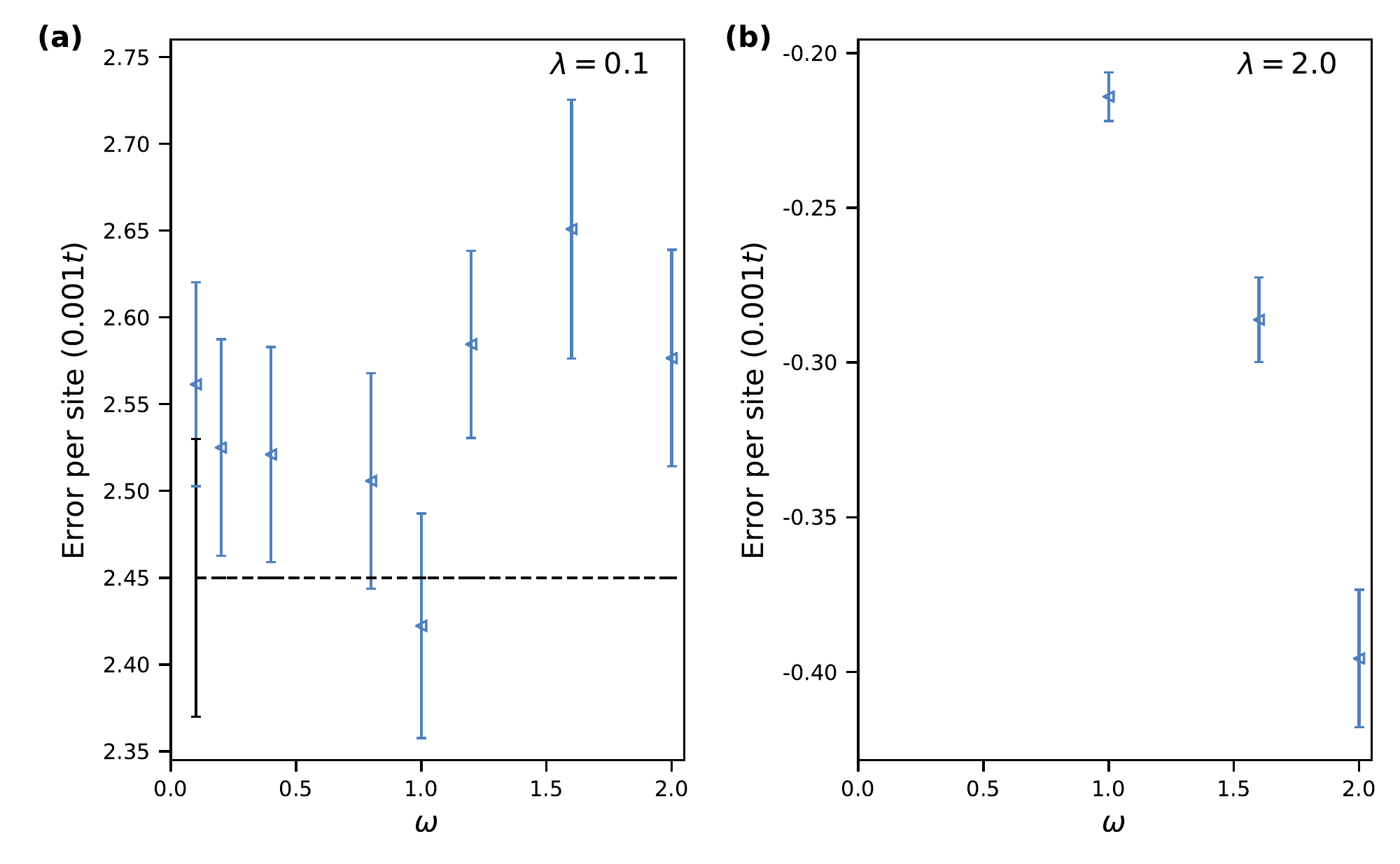}
\caption{\label{fig:20site_error}
Error in total energy per site compared to DMRG in units of $0.001t$ for the 20-site 1D Hubbard-Holstein model at half-filling with $U/t=4$ as a function of $\omega$ for various $\lambda$ values: (a) $\lambda = 0.1$ results and (b) $\lambda = 1.0$ results.
The black dotted line in (a) indicates the electronic CP-AFQMC energy error (i.e., $\lambda = 0.0$).
In (b), DMRG energies are only available for $\omega = 1.0, 1.5, 2.0$.
Note that for (b) the reference DMRG values are potentially unconverged with a maximum phonon number of 40 
and the remaining time step error of CP-AFQMC is not negligible on the plotted energy scale.
}
\end{figure*}

%We also compare the total energy per site obtained from CP-AFQMC to DMRG results in \cref{fig:20site_hh}. DMRG calculations are performed with a bond dimension of 1000 and with the maximum number of bosons of 40. 
%In \cref{fig:20site_hh} (c) and (d), we observe that CP-AFQMC(SDW) closely follows the DMRG energies from $\lambda = 0.1$ to $\lambda = 1.0$. At $\lambda = 2.0$, the energy obtained from CP-AFQMC(SDW) is significantly higher than that obtained from CP-AFQMC(CDW), and CP-AFQMC(CDW) agrees with DMRG very well at $\lambda = 2.0$. If one uses the SDW trial for $\lambda \le 1.0$ and the CDW trial above $\lambda = 1.0$, we expect CP-AFQMC to produce quantitatively accurate results across the full range of parameters that we study. We also note that variations in the phonon frequency do not change the qualitative conclusions stated above.
%
%It is also interesting that CSMP2 still quantitatively fails at $\lambda = 2.0$ because the CDW trial gives nearly perfect CP-AFQMC energies. Furthermore, the second-order correction in $U$ is only 0.00007 $t$. Nevertheless, the smallest error of CSMP2 energy per site from that of CP-AFQMC at $\lambda = 2.0$ is 0.0087 $t$. This is in stark contrast to the 20-site 1D Holstein model at $\lambda = 2.0$ where CSMP2 was within the error bar of CSMP2 at $\omega = 0.1$. 
We next make more quantitative comparison of the total energy per site obtained from CP-AFQMC and DMRG in \cref{fig:20site_hh}. DMRG calculations are performed with a bond dimension of 1000 and with the maximum number of bosons of 40. 
In \cref{fig:20site_hh} (c) and (d), we observe that CP-AFQMC/SDW closely follows the DMRG energies from $\lambda = 0.1$ to $\lambda = 1.0$. At $\lambda = 2.0$, the energy obtained from CP-AFQMC/SDW is significantly higher than that from CP-AFQMC/CDW, with the latter in good agreement with DMRG.
%and CPMC(CDW) agrees with DMRG very well at $\lambda = 2.0$. 
The procedure described above of combining the lowest energy curves between CP-AFQMC/SDW  and CP-AFQMC/CDW
produces quantitatively accurate results across the full range of parameters that we study. 
%If one uses the SDW trial for $\lambda \le 1.0$ and the CDW trial above $\lambda = 1.0$, we expect CPMC to produce quantitatively accurate results across the full range of parameters that we study. We also note that 
Variations in the value of the phonon frequency do not change the qualitative conclusions. % stated above.

 In \cref{fig:20site_error} we show a magnified view of the absolute discrepancies between CP-AFQMC and DMRG energies. 
%We remark that AFQMC energies are not exact for our model even for small $\lambda$ and $\omega$ as shown in \cref{fig:20site_error} (a). Clearly this is due to the constraint bias introduced to remove the sign problem arising from the Hubbard term. 
As a comparison, for the purely electronic Hubbard model 
($\lambda=0$), CP-AFQMC exhibits an error per site of 0.00245(8) $t$ with respect %when compared 
to the DMRG reference values.  Similarly in the %more complex 
Hubbard-Holstein model, CP-AFQMC/SDW energies exhibit an error per site of approximately 0.002--0.003 $t$ for $\lambda = 0.1, 0.3, 0.8$. 
%when compared to DMRG. 
%as in \cref{fig:20site_hh} (a) and (b). 
When the system reaches values as large as $\lambda = 1.0$, we observe a small increase in the error as $\omega$ increases.
% and such an increase becomes comparable to the constraint bias for $\omega > t$. 
The largest error found for $\lambda = 1.0$ is 0.0064(2) $t$ at $\omega = 2.0$, which is slightly larger than the constraint bias found in the purely electronic problem.
\insertnew{The point at which the largest error was observed coincides with the expected phase transition point between SDW and CDW (see below).}
At $\lambda = 2.0$, %the fermionic constraint bias seems to play only a small role and CPMC 
the discrepancy between CP-AFQMC and DMRG is an order of magnitude smaller, 
%in near perfect agreement with DMRG even for the largest $\omega$ examined here. 
%For $\lambda$ values this large, 
with a maximum deviation of CP-AFQMC per site of %is 
-0.00040(2) $t$ at $\omega = 2.0$. 
%Furthermore, the CPMC energies are lower than those of DMRG. At $\lambda = 2.0$ the ground state is mainly dominated by the Holstein physics.  While CPMC is not variational, one may expect that this CPMC energy is ``nearly" variational due to the dominance of the pure Holstein physics. 
Possible reasons for this small discrepancy are that the DMRG energies with a maximum boson number of 40 are not fully converged and the residual time step error in CP-AFQMC is
not negligible on the plotted energy scale.
% however we leave a more careful benchmark study of convergence and comparions between CPMC and DMRG for future work. 

%We comment on the future study of phase diagrams with CP-AFQMC.%
It has been shown by several methods\cite{ohgoe2014variational,Karakuzu2018,Reinhard2019,Costa2020} that in the thermodynamic limit
the Hubbard-Holstein model undergoes a transition between SDW and CDW at
\begin{equation}
U \approx \frac{2g^2}{\omega} = 4\text{d}t \lambda .
\label{eq:crossover}
\end{equation}
This value of $U$ is where the effective on-site interaction changes sign as shown in \cref{eq:lfham}.
For $U/t$ = 4 and $\text{d}=1$, we expect the phase transition to occur at approximately $\lambda = 1.0$. 
%It is of interest to see if our CP-AFQMC algorithm captures this for the finite system sizes we study.
%We find that in the 20-site model, despite the expected finite size effects, the onset of the phase transition is captured quite well.
%In particular, the crossover between CP-AFQMC(SDW) and CP-AFQMC(CDW) occurs roughly at $\lambda = 1.0$. 
%The same onset can be observed in our (near-exact) DMRG calculations for this model as well. 
We find that in the 20-site model, despite the expected finite size effects, the onset of the phase transition is captured quite well.
In particular, the crossover between CP-AFQMC/SDW and CP-AFQMC/CDW occurs roughly at $\lambda = 1.0$ in  \cref{fig:20site_hh}. 
%The same onset can be observed in our DMRG calculations for this model as well. 
%
While this is encouraging, detailed phase diagram studies with CP-AFQMC should be carried out in the future.  
%In this regard, it is important to note that throughout the imaginary time propagation, CPMC often
We note that CP-AFQMC often 
restores the symmetry breaking of the underlying mean-field trial wave function, \cite{Purwanto2008,lee2020utilizing} 
as would be expected of an exact many-body computation. 
%and is able to lead to qualitatively distinct solutions, especially for finite-sized systems.\cite{Purwanto2008,lee2020utilizing}
%In other words, in some cases, CPMC(SDW) does not exhibit the SDW order that its trial wavefunction shows.
Therefore, a proper phase diagram study with CP-AFQMC should involve a direct measurement of correlation functions \cite{chang2010spin}
%that quantify the relevant 
or order parameters  with explicit symmetry-breaking induced.\cite{qin2020absence}
Furthermore, there may be intermediate phases such as metallic or superconducting phases near the onset of the phase transition between the SDW and CDW phases. Studying these putative intermediate phases is of great interest.\cite{Costa2020}
%While encouraging, detailed phase diagram studies with CP-AFQMC should be carried out in the future.  In this regard, it is important to note that throughout the imaginary time propagation, CP-AFQMC often
%restores the symmetry breaking of the underlying mean-field trial wavefunction
%and is able to lead to qualitatively distinct solutions, especially for finite-sized systems.\cite{Purwanto2008,lee2020utilizing}
%In other words, in some cases, CP-AFQMC(SDW) does not exhibit the SDW order that its trial wavefunction shows.
%Therefore, a proper phase diagram study with CP-AFQMC should involve a direct measurement of correlators that quantify the relevant order parameters.\cite{qin2020absence}
%Furthermore, there may be intermediate phases such as metallic or superconducting phases near the onset of the phase transition between the SDW and CDW phases. Studying these putative intermediate phases is of great interest.\cite{Costa2020}

\subsection{2D 4x4 model at half-filling and 1/8 hole-doping}

\begin{figure*}
\includegraphics[scale=0.75]{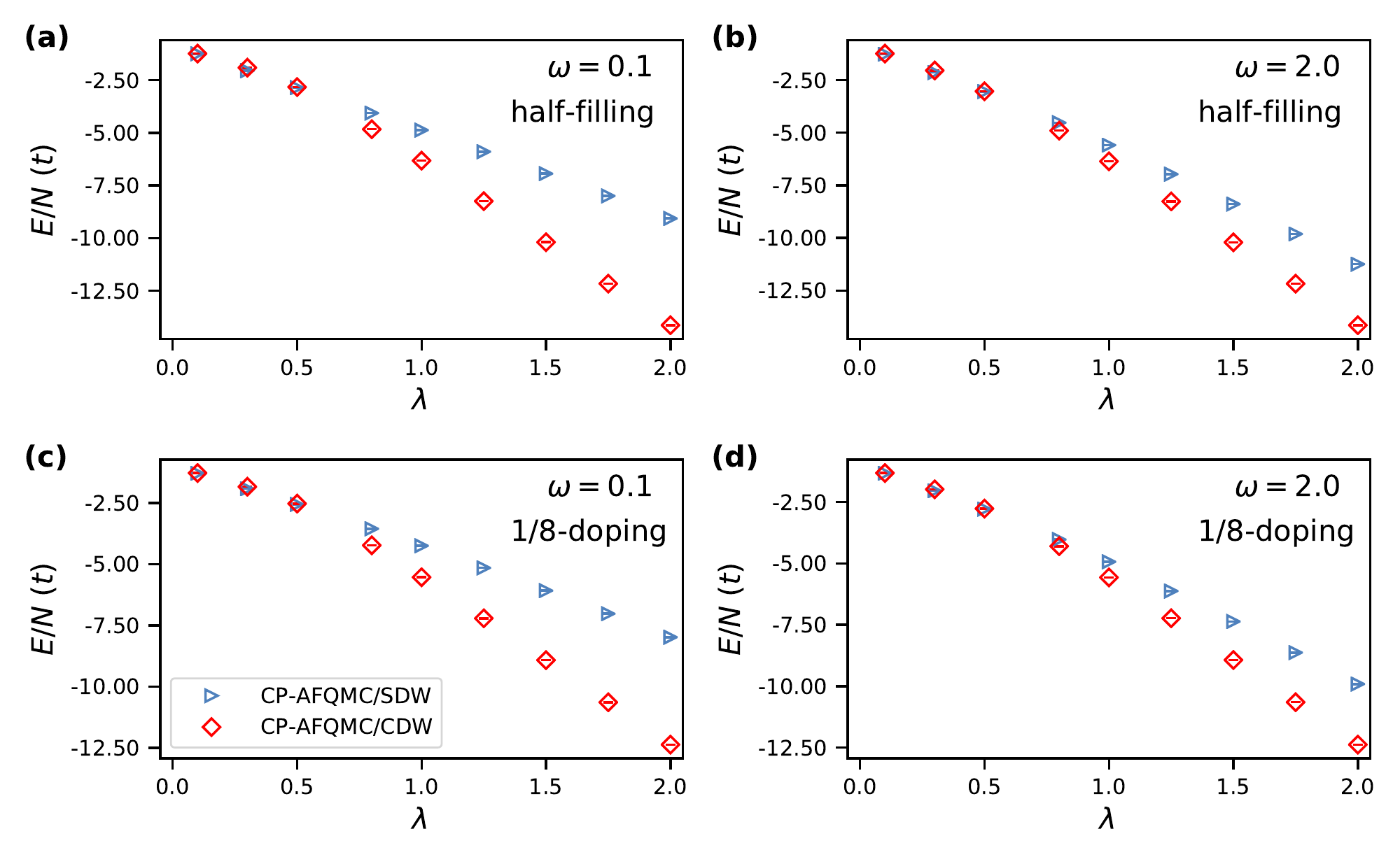}
\caption{\label{fig:2d_hh}
Total energy per site in units of $t$ for the 4x4 2D Hubbard-Holstein model with $U/t=4$ as a function of $\omega$: (a) $\omega = 0.1$ results at half-filling, (b) $\omega = 2.0$ results at half-filling, (c) $\omega = 0.1$ results at 1/8 hole-doping, and (d) $\omega = 2.0$ results at 1/8 hole-doping.
Note that all error bars of CP-AFQMC are too small to be seen on the plotted scales.
}
\end{figure*}

Instead of comparing CP-AFQMC with other methods for 2D Hubbard-Holstein systems, we simply report the computed total energy per site using PBCs along both the $x$ and $y$ directions, as shown in \cref{fig:2d_hh}.
Based on the %careful 
benchmark studies in the previous sections and on experience from the purely electronic model, we expect that our results 
will be of similar accuracy (or better because of effective reduction of the el-el interaction from the el-ph coupling) 
%are at least as accurate as the electronic CPMC calculations are for 
to that in the Hubbard model for most parameters considered in this work.  
In \cref{fig:2d_hh}, we see that  CP-AFQMC/CDW has %trial leads to 
a lower %CPMC 
energy for $\lambda > 0.5$ at both frequencies ($\omega = 0.1$ and $\omega = 2.0$) at both half-filling and 1/8 hole-doping. 
Based on \cref{eq:crossover}, it is expected that the onset of the crossover occurs around $\lambda = 0.5$ in 2D, consistent with our numerical results.
(Note that %This crossover happens at a smaller $\lambda$ value than in %it is in the case of 
%the 1D Hubbard-Holstein model, %simply because 
our definition of $\lambda$ includes dimensionality, hence the change in the crossover value from 1D to 2D).  

Consistent with our previous results, for $\lambda = 0.1$ and $\lambda = 0.3$ at all frequencies up to $\omega = 2.0$, the CP-AFQMC error per site is approximately 0.009 $t$ or slightly larger. % since the underlying trial used is one with SDW order and the Holstein term plays a minor role. 
%\REMARKS{Is the following an empirical observation from our data?}
At the onset of the crossover between CP-AFQMC/CDW and CP-AFQMC/SDW ($\lambda \sim 0.5$), we expect the error to be maximized and larger than that of the CP-AFQMC bias for the electronic problem, \insertnew{similar to the 1D 20-site model at half-filling}.
For $\lambda = 0.8$ and $\lambda = 2.0$, we expect that our results will be nearly exact for the $\omega$ values studied. 
For the purely electronic Hubbard model ($\lambda=0$), CP-AFQMC exhibits an error per site of about 0.00901(9) $t$ at half-filling and 0.00469(4) $t$ at 1/8 hole-doping, using a UHF trial wave function.
As mentioned, %We note that 
the error %in the repulsive Hubbard model 
at half-filling is ``artificial'' and can be removed, %because it can be removed if we used the Hirsch charge decomposition instead of the Hirsch spin decomposition. 
\cite{Shi2013,Qin2016} but this is not done here. 
%Instead of removing this artificial bias, we performed the CPMC calculation as described in this work.
%Therefore, f
For $\lambda \le 1$ at half-filling we expect an error of comparable size.
Comparing two different fillings, we do not see qualitative differences in physical behavior in our finite-sized lattice, and the value of the phonon frequency does not appear to make qualitative differences as well. \insertnew{We note that the energy difference between CP-AFQMC/SDW and CP-AFQMC/CDW noticeably shrinks as the phonon frequency $\omega$ increases.}

\section{Towards {\it ab-initio} Hamiltonians}\label{sec:abinitio}
We briefly discuss the extension of the presented algorithm for
%completely 
general {\it ab-initio} Hamiltonians. 
The {\it ab-initio} Hamiltonian that describes el-ph problems
typically involves linear el-ph coupling.
Therefore, the most widely used {\it ab-initio} Hamiltonian has the 
same form as \cref{eq:HH} with more general  Hamiltonian matrix elements, %complex Hamiltonian energetic and coupling terms,
\begin{align}
\hat{\mathcal H}_\text{el}^\text{(1)} &= 
\sum_{\sigma\in\{\uparrow,\downarrow\}}\sum_{pq}
h_{p_\sigma q_\sigma}
\hat{a}_{p_\sigma}^\dagger
\hat{a}_{q_\sigma},\\
\hat{\mathcal H}_\text{el}^\text{(2)} &= 
\frac12
\sum_{\sigma,\sigma'\in\{\uparrow,\downarrow\}}
\sum_{pqrs}
\left(
p_\sigma r_\sigma 
|
q_{\sigma'} s_{\sigma '}
\right)
\hat{a}_{p_\sigma}^\dagger
\hat{a}_{q_{\sigma'}}^\dagger
\hat{a}_{s_{\sigma'}}
\hat{a}_{r_\sigma},
\\
\hat{\mathcal H}_\text{ph}
&=
\sum_I \omega_I \hat{b}_I^\dagger \hat{b}_I,\\
\end{align}
and
\begin{align}
\hat{\mathcal H}_\text{el-ph}
&=
\sum_{\sigma\in\{\uparrow,\downarrow\}}\sum_{pqI} g_{p_\sigma q_\sigma I} \hat{a}_{p_\sigma}^\dagger
\hat{a}_{q_\sigma}
(\hat{b}_I^\dagger  + \hat{b}_I),
%- g \sum_i \hat{n}_i (\hat{b}_i + \hat{b}_i^\dagger)
\end{align}
where  we have suppressed other quantum numbers such as $\mathbf k$-point dependencies and have expressed everything in terms of the
electronic ($\{p,q,r,s,\cdot\cdot\cdot\}$) and phononic bands($\{I,J,K,L,\cdot\cdot\cdot\}$).
The computation of these matrix elements at the level of density functional theory has been well-documented \cite{giustino2017electron,PONCE2016116,perturbo} so here we focus on briefly describing the {\it phaseless} AFQMC (ph-AFQMC) algorithm\cite{zhang2003quantum} for these realistic el-ph problems.
%\subsection{Propagation}

The walkers take the same form as in \cref{eq:walkers}. Therefore, the essence of the propagation algorithm remains unchanged.
The only complication arises from the generalized form of $\hat{\mathcal H}_\text{el}^\text{(2)}$ which necessitates the use of a continuous Hubbard-Stratonovich transformation.\cite{Stratonovich1957,hubbard1959calculation} The continuous transformation leads to the fermionic phase problem which can be removed %completely 
via the phaseless constraint. \cite{zhang2003quantum} The propagation is carried out the same way with appropriate modifications to the constraint to account for the phase problem.
The {\it ab-initio} generalization of semiclassical states used in this work is also straightforward.
The trial wavefunction still takes the form of \cref{eq:cs}, \cref{eq:csel}, and \cref{eq:csph}.
A variational minimization of the total energy then leads to a trial wavefunction that can be used in ph-AFQMC.
The projection of the trial wavefunction onto phonon displacements $|\mathbf X\rangle$ is identical to \cref{eq:trialprojection}
except that the phonon mass and frequency now depend on band indices $\{I\}$. 
\insertnew{The {\it ab-initio} generalization of the LF wavefunction may also be carried out straightforwardly by extending the LF generator in \cref{eq:lfU}.}
We expect that the {\it ab-initio} ph-AFQMC approach will become a valuable tool for understanding polaronic physics in realistic correlated materials in the future.

\section {Conclusions and outlook}\label{sec:conclusions}
In this work,
we have introduced an
extension of CP-AFQMC %constrained-path Monte Carlo (CP-AFQMC)
to describe
correlated systems with el-ph coupling.
Our approach utilizes a mixed first/second-quantized representation
where the phonons are propagated in first quantization following the commonly used diffusion MC algorithm, and the
electronic degrees of freedom are handled in second quantization via AFQMC.
%CP-AFQMC which decouples the electron-electron interaction term with the Hirsch spin decomposition.\cite{Hirsch1983}
The resulting algorithm is
compared with numerically-exact DMRG %density matrix renormalization group theory (DMRG)
and low-order perturbation theories for the Holstein model as a first test of the basic algorithm.
We have demonstrated that the autocorrelation time problems that arises in the commonly used DQMC methods is greatly ameliorated in AFQMC, with autocorrelation time that scales roughly as $1/\omega$. % as opposed to the $1/\omega^2$ scaling in DMC.

%\REMARKS{11/9: leaving the following here but suggest we remove it pending any contradiction from current tests:}
%\REMARKS{where does the following stand now? I believe the answer should be exact in this situation. 
%The calculation can be terribly biased, in the same sense as seen in the 2-site problem, when the importance function is poor (and aggressive).
%The data here is much less clear that there's a systematic discrepancy with DMRG. If there is one where this is clear, can we check with a softer 
%importance function?  }
\insertnew{While the Holstein model is sign-problem free, AFQMC with the simplest trial wavefunctions, namely semiclassical states,
is found to introduce a small bias when the underlying adiabatic surface bifurcates and/or $g$ is larger than $t$ but smaller than $\omega$ (e.g., $\omega \rightarrow \infty$ for a fixed $\lambda$).
%We have suggested that this bias is due to an ergodicity problem associated with the structure of these states, and leave a more thorough investigation of this for future work. 
Based on a 4-site model, we have shown that this bias can be removed
by using an improved trial wavefunction where both bifurcations and increased el-ph correlation are accounted for. 
%As an example, we showed that a linear combination of linearized Lang-Firsov states is enough to unbias the result in the case of the 4-site problem.
%We have also demonstrated that the autocorrelation time problems that arises in the commonly used DQMC method is greatly ameliorated, with a AFQMC autocorrelation time that scales roughly as $1/\omega$ as opposed to the $1/\omega^2$ scaling in DQMC.
We have demonstrated the remarkable accuracy of AFQMC for both 1D and 2D Holstein models over a reasonably broad set of coupling and phonon frequency parameters via direct comparison with DMRG.}

We have tested CP-AFQMC on the finite sized versions of the 1D and 2D Hubbard-Holstein models with $U/t=4$, using the simplest form of trial wave functions consisting of a semiclassical state with a single Slater determinant. 
For the 1D Hubbard-Holstein model, we have compared CP-AFQMC against numerically exact DMRG results.
When $\lambda$ is small and the ground state is dominated by the Hubbard $U$ term, we find that
the error of our algorithm is roughly the same as that expected from standard CP-AFQMC applied to the purely electronic Hubbard model. Furthermore, when the ground state is dominated by the el-ph coupling term and exhibits charge density wave order, 
we find that the overall error becomes remarkably small (smaller than that expected in purely electronic systems).  These facts have motivated the production of what we believe are benchmark results for the finite sized 2D Hubbard-Holstein model for various values of $\lambda$ and $\omega$ at half-filling and 1/8 hole-doping. 

For $U<2g^2/\omega$ at half-filling and for all parameter regimes at \insertnew{any} hole-doping, standard QMC approaches suffer from the sign problem.\cite{Karakuzu2018} Therefore, our AFQMC approach should become an essential tool for producing accurate results scalable to large system sizes for this model. 
We have %also discussed two possible trial wavefunctions for the 
investigated the competing spin- and charge-density wave orders in the 
Hubbard-Holstein model. % namely one with charge density wave order and one with spin density wave order. 
At the onset of the phase transition between these phases, we observe a crossover in the energies between two AFQMC calculations targeting the two 
phases. %energies based on these trials, signaling that trail choice may be easily dictated by the energetics of the system. 
Lastly, we have briefly discussed the extension of the presented algorithm to {\it ab-initio} Hamiltonians that can be easily formulated based on 
the phaseless AFQMC method for general electronic Hamiltonians. \cite{zhang2003quantum} 

Some immediate future directions include using this framework to provide a detailed study of the phase diagram of the Hubbard-Holstein model and other lattice models, and
extending this framework to finite-temperature problems based on the constrained path approximation.\cite{Zhang1999}
\insertnew{
%A direct, efficient implementation of the full Lang-Firsov trial wavefunction-based AFQMC approach outlined here will be an important research direction which will help provide unbiased energy estimates for anti-adiabatic regimes with $2\text{d}t\lambda < \omega$.
As mentioned, a trial wave function with an el-ph Jastrow factor can be implemented 
straightforwardly in AFQMC, which is expected to  
further reduce the bias and improve sampling efficiency in anti-adiabatic regimes with $2\text{d}t\lambda < \omega$. 
It will also be valuable to further investigate the implementation of 
the full LF trial wave function.
} 
Furthermore, application of the proposed AFQMC approach to {\it ab-initio} systems will be of great interest as well.
While there are several algorithmic aspects that can be further improved, including improved forms of importance functions and better sampling 
in large phonon frequency regimes,
%in highly multi-modal parameter regimes, 
we believe that the algorithms and insights presented in this work will serve as stepping stones towards
simulating model as well as {\it ab-initio} systems with a non-trivial interplay between electronic correlation and el-ph couplings, which continue to be of great importance in modern condensed matter physics.

\section{Computational Details}
\label{sec:comp-details}

Our  algorithm was implemented in a public open-source auxiliary-field quantum Monte Carlo package called \texttt{PAUXY}.\cite{pauxy}
The blocking analysis was performed with \texttt{pyblock}.\cite{pyblock} The pair branching algorithm was used for population control.\cite{wagner_qwalk}
Variational calculations were aided by automatic differentiation using \texttt{JAX}.\cite{jax2018github}
A total of 640 walkers and a time step of 0.005 $t^{-1}$ were used in all calculations except for the Holstein dimer \insertnew{and the 4-site 1D Holstein model}. 
For the Holstein dimer, we used a time step of 0.0005 $t^{-1}$ for $\omega \le 1.6$ and 0.00025 $t^{-1}$ for $\omega > 1.6$ \insertnew{with 6400 walkers}. 
\insertnew{For the 4-site 1D Holstein model, we used 6400 walkers and a time step of 0.0005 $t^{-1}$.}
The population control bias and time step error were found to be smaller than 0.001 $t$ in the absolute total energy per site.
%\insertnew{We note that we do not consider the sampling bias in the Holstein model to be the population control bias here.}
\section{Acknowledgement}
We are grateful to Zihang Li and Hao Shi for their contributions during early stages of this work. We thank
Hao Shi, Michael Lindsey, Fionn Malone for fruitful discussions and Miles Stoudenmire for help with iTensor calculations. DRR acknowledges support from grant NSF-CHE 1954791.
This work was conducted
using computational resources and services at the Flatiron Institute. The Flatiron Institute is a division of the Simons Foundation. 
\bibliography{main}
\end{document}